\documentclass[%
 reprint,
superscriptaddress,
 amsmath,
 amssymb,
 aps,
prb,
]{revtex4-2}
\usepackage{graphicx}% Include figure files
\usepackage{dcolumn}% Align table columns on decimal point
\usepackage{bm}% bold math
\usepackage[utf8]{inputenc}
\usepackage[T1]{fontenc}
\usepackage{mathptmx}
\usepackage{etoolbox}
\usepackage{xcolor}
\usepackage{upgreek}
\usepackage{physics}
\usepackage{hhline}

\usepackage{blkarray}
\usepackage{placeins} 

\newcommand{\eran}[1]{{\color{black}#1}}
\newcommand{\akm}[1]{{\color{black}#1}}

\makeatletter
\def\@email#1#2{%
 \endgroup
 \patchcmd{\titleblock@produce}
  {\frontmatter@RRAPformat}
  {\frontmatter@RRAPformat{\produce@RRAP{#1\href{mailto:#2}{#2}}}\frontmatter@RRAPformat}
  {}{}
}%

\makeatother
\begin{document}

% \preprint{AIP/123-QED}

\title{Hybrid metal-semiconductor quantum dots in InAs as a platform for quantum simulation}
% Force line breaks with \\
\author{Praveen Sriram}
\thanks{These authors contributed equally to this work.}
\author{Connie L. Hsueh}
\thanks{These authors contributed equally to this work.}
\affiliation{
Department of Applied Physics, Stanford University, Stanford, CA, USA}
\affiliation{
Stanford Institute for Materials and Energy Sciences, SLAC National Accelerator Laboratory, Menlo Park, CA, USA}
\affiliation {Geballe Laboratory for Advanced Materials, Stanford University, Stanford, CA, USA}

\author{Karna A. Morey}
\affiliation {Geballe Laboratory for Advanced Materials, Stanford University, Stanford, CA, USA}
\affiliation{
Department of Physics, Stanford University, Stanford, CA, USA}

\author{Tiantian Wang}
\affiliation{
Department of Physics and Astronomy, Purdue University, West Lafayette, IN, USA}
\affiliation{Birck Nanotechnology Center, Purdue University, West Lafayette, IN, USA}

\author{Candice Thomas}
\affiliation{
Department of Physics and Astronomy, Purdue University, West Lafayette, IN, USA}
\affiliation{Birck Nanotechnology Center, Purdue University, West Lafayette, IN, USA}

\author{Geoffrey C. Gardner}
\affiliation{Birck Nanotechnology Center, Purdue University, West Lafayette, IN, USA}
\affiliation{
Microsoft Quantum Lab West Lafayette, West Lafayette, IN, USA}

\author{Marc A. Kastner}
\affiliation{
Stanford Institute for Materials and Energy Sciences, SLAC National Accelerator Laboratory, Menlo Park, CA, USA}
\affiliation{
Department of Physics, Stanford University, Stanford, CA, USA}
\affiliation{
Department of Physics, Massachusetts Institute of Technology, Cambridge, MA, USA}

\author{Michael J. Manfra}
\affiliation{
Department of Physics and Astronomy, Purdue University, West Lafayette, IN, USA}
\affiliation{Birck Nanotechnology Center, Purdue University, West Lafayette, IN, USA}
\affiliation{
Microsoft Quantum Lab West Lafayette, West Lafayette, IN, USA}
\affiliation{Elmore Family School of Electrical and Computer Engineering, Purdue University, West Lafayette, IN, USA}
\affiliation{School of Materials Engineering, Purdue University, West Lafayette, Indiana 47907, USA}

\author{Andrew K. Mitchell}
\affiliation{
School of Physics, University College Dublin, Belfield, Dublin 4, Ireland
}
\affiliation{
Centre for Quantum Engineering, Science, and Technology, University College Dublin, Dublin, Ireland}

\author{Eran Sela}
\affiliation{
Raymond and Beverly Sackler School of Physics and Astronomy, Tel Aviv University, Tel Aviv, Israel
}

\author{David Goldhaber-Gordon}
% \thanks{Corresponding author}
\email{Corresponding author : goldhaber-gordon@stanford.edu}
\affiliation{
Stanford Institute for Materials and Energy Sciences, SLAC National Accelerator Laboratory, Menlo Park, CA, USA}
\affiliation{
Department of Physics, Stanford University, Stanford, CA, USA}

\date{July 27, 2026}
\begin{abstract}
\section*{Abstract}
Arrays of hybrid metal-semiconductor quantum dots offer a new approach to quantum simulation, with key advantages over arrays of conventional quantum dots. Because the metallic component of these hybrid dots has a quasi-continuous level spectrum, each site in an array can be effectively electronically identical; in contrast, each conventional semiconductor quantum dot has its own spectral fingerprint. Meanwhile, the semiconductor component retains gate-tunability of intersite coupling. This combination creates a scalable platform for simulating correlated ground states driven by Coulomb interactions. We report the fabrication and characterization of hybrid metal-semiconductor dots, featuring a submicron metal island transparently contacting a gate-confined region of an InAs quantum well with tunable couplings to macroscopic leads. Tuning the dot-lead coupling to the weak-coupling limit yields highly-uniform Coulomb peaks, with no resolvable excitation spectrum in the Coulomb diamonds. We propose a realistic device design for a hybrid dot-based linear array quantum simulator and outline the correlated many-body physics accessible in this architecture.
\end{abstract}
\newcommand{\Gset}[0]{G_{\text{SET}}}
\maketitle
\section*{Introduction}
Our best way to understand correlated electron behavior in an exotic material is to try to describe that material through a simplified effective Hamiltonian. Yet even that simplified Hamiltonian is often too computationally intensive to solve in detail. This has spurred efforts to build tunable experimental realizations of such Hamiltonians, to measure their properties instead of calculating them on a conventional classical computer. One prominent approach to realizing such Hamiltonians is based on arrays of a few semiconductor quantum dots~\cite{VanDiepen2021,Hsiao2024,Kim2022,Knrzer2022}. The strong Coulombic onsite energy $U$, electrostatic tunability of tunnel couplings $\mathcal{T}_{\text{tunnel}}$, and low carrier temperature $T_e \ll U/k_B$, where $k_B$ is the Boltzmann constant, provide access to a parameter space favorable for studying strongly-interacting ground states and quantum phase transitions among them. Exotic phases and critical states simulated in this way include the Mott insulator~\cite{Vandersypen2017}, resonating valence bonds~\cite{Wang2023}, Nagaoka ferromagnetism~\cite{Dehollain2020}, and single- and two-channel Kondo critical points~\cite{goldhaber-gordon_kondo_1998, vanderWiel2000, Potok2007}. To connect to bulk material properties, scaling up to larger arrays will be important. \textcolor{black}{More generally, realizing a quantum dot-based quantum simulator requires (i) at each site in the array, a gapped spectrum with a well-defined and gate voltage-tunable ground state, (ii) tunable tunnel coupling between each pair of neighboring sites, and (iii) a low effective temperature compared to the on-site gap and other relevant energy scales. Specific applications may impose more stringent requirements on site-to-site variations in the charging energy, excited state spectra, and the range of tunability of the inter-dot tunnel couplings~\cite{Barthelemy2013}}.

Progress is being made on many technical challenges to such scaling, including gate fanout~\cite{Nguyen2017}, cryostat wiring and control interface limitations~\cite{DJReilly2019}, and crosstalk between control of different dots in the array~\cite{Undseth2023}. Borsoi {\em et al.}~\cite{Borsoi2022} demonstrated shared gate control of a 4$\times$4 array of dots based on holes in Ge, tuning all 16 dots simultaneously to single-electron occupancy. However, charging energies still varied by 10-20\% across the array. 
Even with impressive advances from leveraging industrial wafer-scale fabrication processes~\cite{Neyens2024}, such dot-to-dot inhomogeneity remains a key obstacle to realizing the large arrays with essentially identical sites necessary for quantum simulation of models relevant to bulk material properties. \textcolor{black}{Among the} most problematic are site-to-site variations in wavefunctions and thus excited-state spectra and inter-dot tunnel couplings. 

\begin{figure}[!htbp] % or [h], [b], etc.
    \centering
    \includegraphics[width=0.49\textwidth]{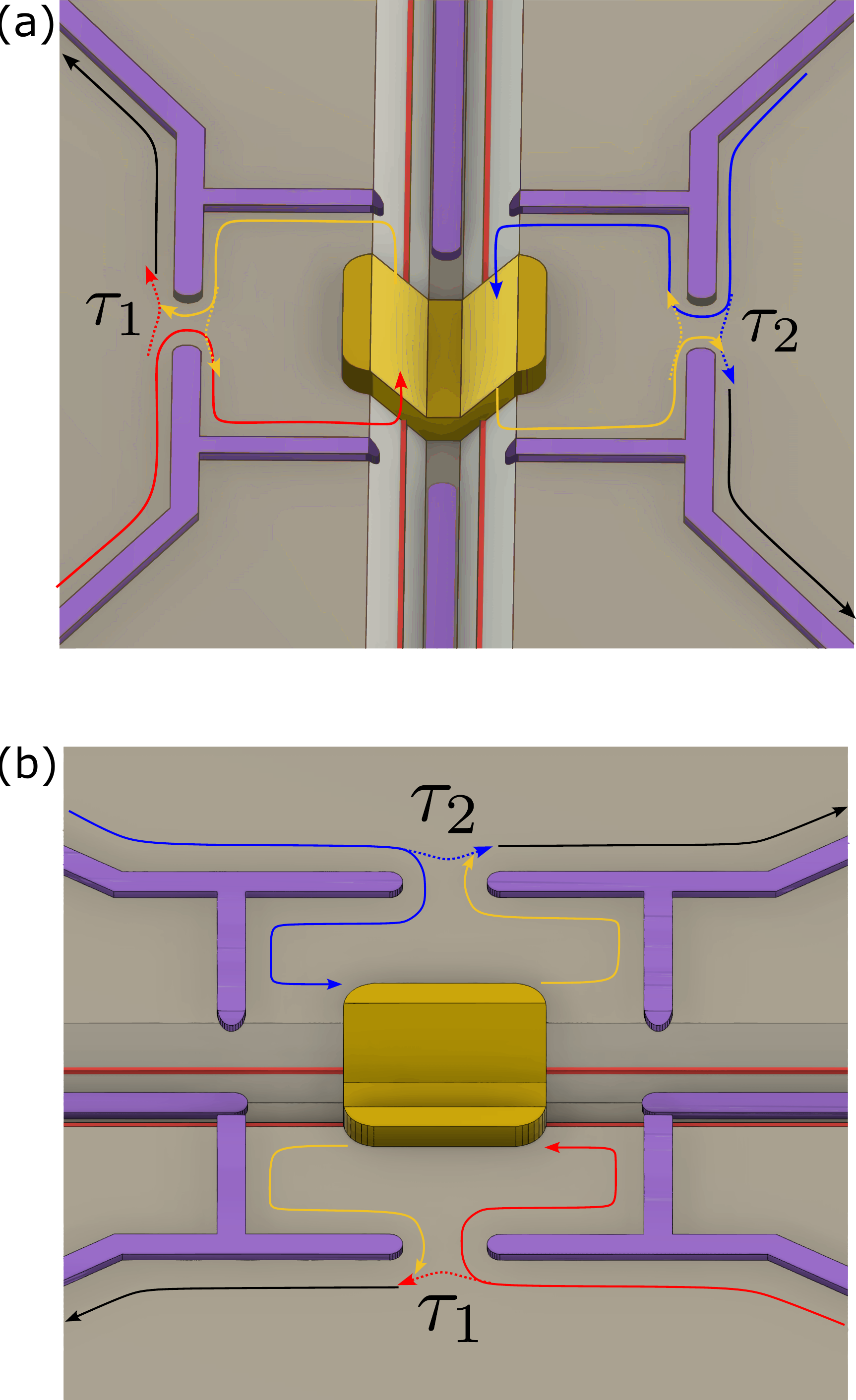}
    \caption{\textbf{Device design schematic.} Schematic representation of the hybrid metal-semiconductor dot architecture in (a) view along the axis of the wet-etched trench, and (b) perpendicular view consistent with later descriptions and figures. The InAs two-dimensional electron gas (red layer within the gray semiconductor heterostructure stack) is confined with gate electrodes (purple) that realize a pair of quantum point contacts to tune the tunnel coupling to quantum Hall edge modes (red, blue) with transmissions $\tau_{1,2}$. A submicron metallic island strongly couples to and equilibrates the incident edge modes, resulting in a metallic density of states in the confined region and downstream (yellow) edge modes at the island chemical potential.}
    \label{fig:example}
\end{figure}

A recently-introduced platform based on hybrid metal-semiconductor dots~\cite{Iftikhar2015, Pouse2023} \textcolor{black}{offers a promising route to address} this problem of dot-to-dot inhomogeneity. Each individual site consists of a small metal island electrically connected to a comparably-sized gate-confined region of two-dimensional electron gas (2DEG). Each site is tunably coupled to reservoirs~\cite{Iftikhar2015} and/or neighboring sites by quantum point contacts. A schematic of the device design is shown in Fig.~\ref{fig:example}. The energy spectrum of the confined region is characterized by a charging energy $E_c = e^2/2C$, where 
$C$ is the total capacitance of the hybrid dot, determined by the geometry of the island and gate electrodes as well as the dielectric environment. The metallic component grants the hybrid dot a continuous density of states \textcolor{black}{that is expected to be largely} insensitive to the confinement potential landscape, \textcolor{black}{improving uniformity across sites as compared with conventional semiconductor quantum dots~\cite{Iftikhar2015}}. The semiconductor component enables tuning intersite and site-reservoir coupling. As a refinement, a high magnetic field is applied to put the 2DEG in an integer quantum Hall state, bringing two advantages. First, in the quantum Hall regime reflection from a quantum point contact is easy to measure. Since every electron must either transmit or reflect, measuring reflection also directly quantifies transmission. Second, reduced backscattering from some types of potential disorder can make transmission a smoother function of gate voltages than at zero field \cite{vanweesQuantizedConductanceMagnetoelectric1988,vanweesQuantumBallisticAdiabatic1991, vanhoutenChapter2Quantum1992}. Recent experiments on a pair of hybrid metal-GaAs quantum dots~\cite{Pouse2023} investigated a non-Fermi liquid quantum critical point (QCP) between competing ‘charge-Kondo'~\cite{Pierre3CK, Iftikhar2015, piquard2026experimental} ground states. Scaling this to many such sites \textcolor{black}{offers a promising route} to simulating Kondo lattice models and investigating lattice coherence in heavy-fermion materials.

For a hybrid dot to act as a single charging site, contact between the metal and semiconductor components must be highly transparent~\cite{Iftikhar2015}. In previous implementations using a 2DEG in a GaAs-based heterostructure as the semiconductor, this has been a formidable challenge: the midgap pinning of the Fermi level at the interface between metal and GaAs~\cite{woodall_ohmic_1981} requires the metal component in Refs.~\cite{Pouse2023, Pierre3CK, Iftikhar2015} to be a few microns wide to achieve reliable contact with transparency $>99$\%. This in turn makes a site's charging energy small ($300$~mK $k_B$ \textcolor{black}{$\approx 26$ $\upmu$eV}), limiting measurements of quantum critical phenomena to a narrow temperature range bounded from above by one quarter of the charging energy ($\approx 70$~mK) and from below by the lowest achievable electron temperatures in such structures ($\approx 10-20$~mK). Increasing the charging energy would open a larger range of measurement temperatures, allowing far more detailed quantitative analyses of quantum critical behavior \cite{Mitchell2016}. \textcolor{black}{In summary, simulating charge-Kondo lattice models in the hybrid metal–semiconductor platform shares some requirements with more conventional quantum dot-based quantum simulation: large charging energies with minimal site-to-site inhomogeneity, and smoothly tunable QPCs that allow control of tunnel couplings. However, it also has distinctive added requirements: highly transparent metal–semiconductor interfaces, and extended tunability of QPCs all the way from the weak tunneling to the near-ballistic regime. Operating in the quantum Hall regime brings important advantages, and this requires a well-defined quantum Hall state in the semiconductor 2DEG.}

InAs offers the prospect of higher charging energies based on more miniaturized ohmic contact to the semiconductor. Fermi level pinning above the conduction band minimum instead of in the gap~\cite{woodall_ohmic_1981} obviates the need for annealed eutectic Au/Ge alloy contacts, a major source of inhomogeneous contact morphology~\cite{Goktas2009}. We report the fabrication and characterization of \textcolor{black}{a single-site building block of a} hybrid metal-InAs quantum dot\textcolor{black}{-based architecture}, featuring a $\approx 1$ $\upmu$m$^2$ island with tunable couplings to macroscopic reservoirs. In the quantum Hall regime, a near-perfect interface transparency (>99.2\%) is measured for the outermost edge mode at quantum Hall filling factor $\nu = 2$. Quantum point contacts (QPCs) provide electric control of the dot-reservoir couplings. At weak tunneling we measure charge quantization with strong uniformity in Coulomb peaks, no resolvable excitation spectrum in the Coulomb diamonds (Fig~\ref{fig:CB}), and a four-fold enhancement in the charging energy over hybrid GaAs-based dots reported by us and others. Using QPCs nominally identical to those in Ref.~~\cite{Hsueh2022}, we tuned the coupling strengths to reveal the evolution from static to dynamical Coulomb blockade, with charge quantization fully quenched when QPCs are opened to transmit a full mode. Our work positions hybrid metal-InAs quantum dots as a \textcolor{black}{promising} candidate platform for making arrays of functionally-identical sites to explore Kondo lattice coherence.

% \section{Device Fabrication \& Measurement setup}
The devices featured in this work were fabricated on a heterostructure grown by molecular beam epitaxy on a semi-insulating InP (100) substrate. A 4 nm InAs quantum well sandwiched between 10.5 nm In$_{0.75}$Ga$_{0.25}$As layers forms the 2D channel. The InAs quantum well is populated by unintentional impurities from deep-level donor states in the In$_{0.75}$Al$_{0.25}$As layers~\cite{Luo1993,Capotondi2004}. 2DEG density $n=4.8 \times 10^{11}$ cm$^{-2}$ and mobility $6\times 10^5$ cm$^2$/Vs are measured in a 60-$\upmu$m-wide Hall-bar at 40~mK (see Methods). This mobility is enhanced by aspects of the heterostructure design: a 900 nm linearly graded In$_{x}$Al$_{1-x}$As strain-relaxed buffer layer, and a 120 nm In$_{0.75}$Al$_{0.25}$As capping layer. 

The devices were processed with electron beam lithography and a citric/phosphoric acid-based wet etch to define the mesas (see Methods for details). Aiming for a pristine metal-InAs interface on the submicron-scale, we avoided oxygen plasma or $in$ $situ$ Ar ion milling in the contact region, instead depositing Ti/Au to make edge contact to the quantum well immediately after a short HCl dip. After contact deposition, a blanket 35 nm HfO$_2$ gate dielectric layer was grown by atomic layer deposition at 150$^{\circ}$C, followed by deposition and liftoff of patterned Ti/Au gate electrodes. 

The measurements reported here were performed in a dry He$^3$-He$^4$ dilution refrigerator with a mixing base temperature T$_{0} = 34 $~mK, over multiple cooldowns. Unless otherwise specified, an out-of-plane magnetic field $B_{\perp} = 5.5$ T placed the 2DEG in the integer quantum Hall phase with filling factor $\nu = 3$ and counter-clockwise chirality. The vanishing longitudinal resistance, quantized Hall resistance, and chirality of current flow simplify the analysis of transport data, permitting a quantitative treatment of the conductance as a function of arbitrary tunnel couplings at the point contacts. With Zeeman effect lifting spin degeneracy, each of the two QPCs can couple the dot to a reservoir through a spin-polarized edge channel. As we have just noted, quantum Hall edge modes are used because chiral current flow provides an operational advantage for transport characterization. The implications of chiral current flow for lattice Hamiltonians in the context of quantum simulation are discussed in the corresponding Results section.

Transport is measured by sourcing an ac current excitation $I_{ac} = 500$~pA - 20~nA through a 100~M$\Omega$ bias resistor, at frequencies below 100~Hz. The current is drained at both ends of the InAs mesa through grounded contacts tied directly to the cold finger of the dilution refrigerator. The robust and broad quantum Hall plateau converts the source current into an on-chip differential voltage excitation $V_{ac} = I_{ac} \times h/\nu e^2$, where $h/e^2$ is the resistance quantum. The voltage relative to ground is measured at voltage probes downstream of the hybrid quantum dot, using high-impedance voltage preamplifiers and standard lock-in techniques. For a perfectly transparent metal-semiconductor interface, a Landauer-B{\"u}ttiker treatment reveals the series conductance $G$ through the device as 

\begin{equation}
    G = \nu \frac{e^2}{h}\frac{V_T}{V_T + V_R},
    \label{eq:QHbuttikerconductance}
\end{equation}
where ${V_{T, R}}$ denote the transmitted and reflected voltage drops, measured as described above, in the geometry shown in Fig.~\ref{fig:transparency}(a).

Data from two devices are reported in this work. To study transparency of a metal island to quantum Hall edge modes, Device A features a submicron Ti/Au island bridging a trench on a mesa, making edge contact to the InAs quantum well on each side of the trench. Two QPCs allow independent probing of the transparency of each edge mode. The QPCs are placed far from the island (60 $\upmu$m) to suppress charging effects (see the Transparency section). To study charge quantization and charge noise, Device B again features a submicron metal island, but now with QPCs far closer to the island (400 nm) to form a hybrid quantum dot, as demonstrated in the Static Coulomb blockade section.

\section*{Results}
\subsection*{Transparency of the submicron metallic island to quantum Hall edge modes}
\label{sec:Transparency}
\begin{figure*}[!htbp]

     \centering
         \includegraphics[width=1.0\textwidth]{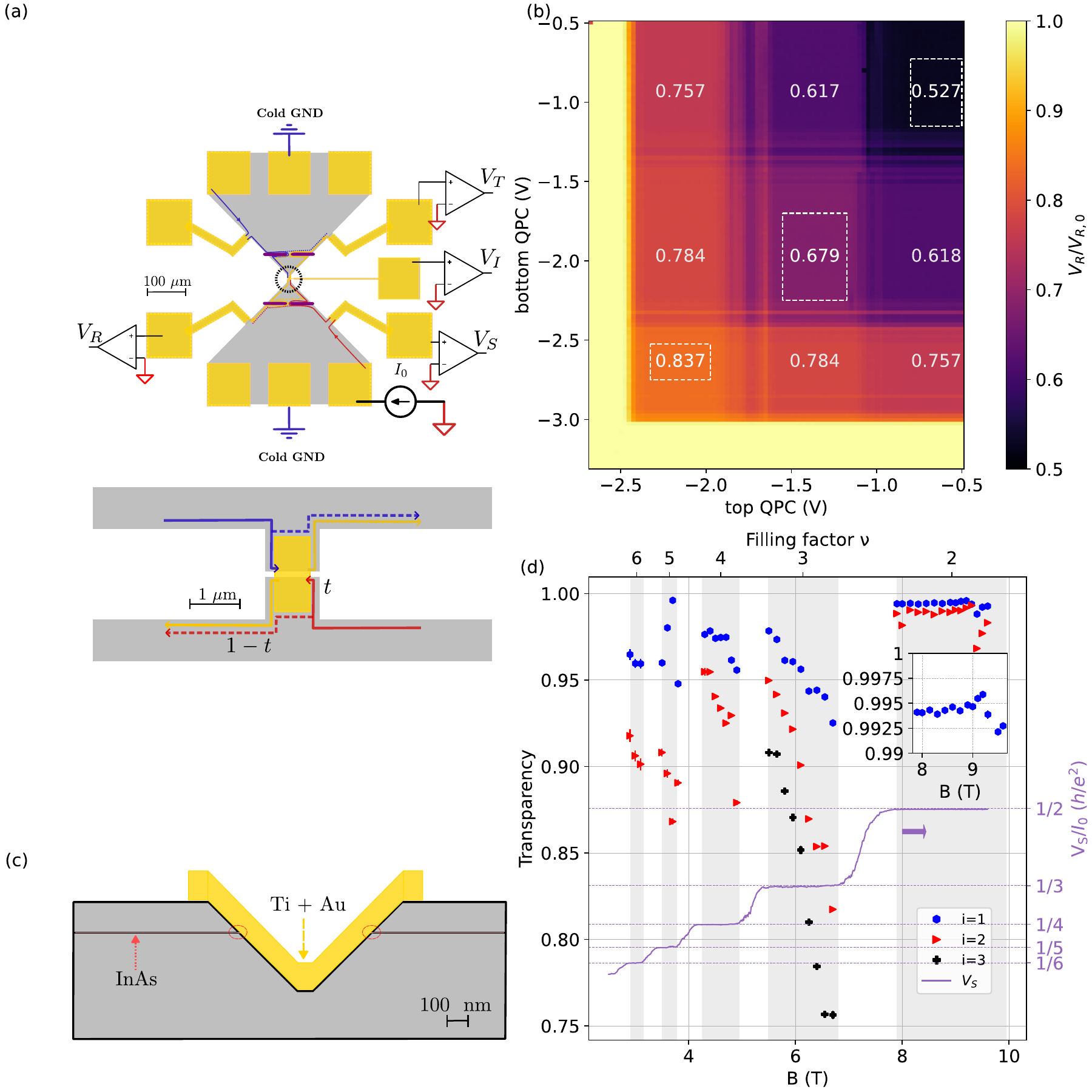}

        \caption{\textbf{Edge state transparency}. (a) Device and measurement setup schematic. Two split-gate QPCs (purple) are each tuned to transmit an integer number of quantum Hall edge modes $i$ into a central region of the InAs mesa (gray), shown expanded in the bottom panel. Edge states are launched from a current-biased Ti/Au ohmic contact (yellow) at lower right, and are collected at a pair of contacts tied to a ``cold ground'' at the mixing chamber stage of the dilution fridge. The source ($V_S$), reflected ($V_R$), and transmitted ($V_T$) voltages are measured $\approx$ 160 $\upmu$m from the central region. Bottom panel: A 0.7$\times$1.5 $\mu$m$^2$ Ti/Au (yellow) island ohmic electrically bridges across a trench in the narrowest, 1.1-$\upmu$m-wide, section of the mesa. Potential on this island ($V_I$) is also measured. 
        Red and blue solid lines denote edge modes transmitted through the QPCs on the source and drain sides respectively. Dashed lines of the corresponding colors denote modes reflected from interfaces with the island: with a probability $(1-t_i)$, the $i$-th outer edge mode is reflected. Yellow lines denote modes emitted from the island. (b) Normalized reflected voltage as a function of the QPC gate voltages at $B= 5.5$ T ($\nu = 3$), clearly showing regions of quantized transport for 1, 2, and 3 transmitted edge modes through each of the two QPCs. Each such region is annotated with the average measured value of the reflected voltage. Collectively, these values allow extracting the transparency of each edge mode's transmission into the metal island. (c) A cross-sectional view of the Ti/Au island making sidewall contact to the InAs quantum well, with the interface marked with red dashed circles. (d) Edge state transparency as a function of field for $2\leq \nu \leq 6$, with the outermost edge mode indexed $i=1$. The inset shows the outermost edge mode transmission $t_1>99.2\%$ at $\nu=2$. }
        \label{fig:transparency}
\end{figure*}

The quality of the ohmic contact made by a metal island is characterized by the reflection probability of the quantum Hall edge modes at the interface between metal and quantum well. Device A for measuring transparency involves a mesa broken by a trench across its narrowest portion. We choose to make the trench 400 nm deep, to etch through not only the 2DEG but also In-rich buffer layers that would otherwise conduct across the trench when exposed by a shallower etch. Though the nominal, lithographically-defined trench width is only 120 nm, the nearly isotropic wet etch process's $\approx$ 45$^{\circ}$ sidewalls make the actual etched region far broader. A Ti/Au island with a $0.7\times 1.5$ $\upmu$m$^2$ footprint is deposited across the trench, making sidewall contact to the exposed InAs quantum well on each side of the trench. Macroscopic ohmic pads serve as the source and grounded-drain terminals, and as voltage probes. This geometry ensures that any current between the two mesa regions must flow through the metal island. Figure~\ref{fig:transparency}(a) shows the device geometry, with the cross-section in Fig.~\ref{fig:transparency}(c) depicting the trench with the metal island making sidewall contact to the quantum well. A pair of QPCs $60$ $\upmu$m on either side of the island are used to selectively populate edge modes, allowing a thorough study of transparency as a function of edge state index. 

For a perfectly transparent interface, the edge modes transmitted through the QPCs will fully equilibrate at the island. Fitting the measured downstream voltages $V_T$, $V_R$ to a Landauer-B{\"u}ttiker model provides an estimate of transparency.

When the QPCs each transmit only the outermost edge mode, the reflected voltage $\left(V_{R}^{\tau_{1,2}=1}\right)$ normalized to its value with the QPCs pinched-off $\left(V_{R}^{\tau_{1,2}=0}\right)$ is given by
\begin{equation}
{V}_{R,1} := \frac{V_R^{\tau_{1,2} = 1}}{V_R^{\tau_{1,2} = 0}} = 1 - \frac{t_1}{2\nu},
\label{eq:transparencyouterLB}
\end{equation}
where $1-t_1$ is the reflection probability of the outer edge mode at the island-quantum well interface, and $\tau_{1,2}$ are the number of edge modes transmitted through the two QPCs. Description of the model and a derivation of this expression are included in the SM. Figure~\ref{fig:transparency}(b) depicts the normalized reflected voltage $V_{R,i}$ as a function of the QPC gate voltages, clearly showing quantized plateaus corresponding to positive integer $i \in \mathbb{Z^+} \leq \nu$ edge modes transmitted through each QPC. From the average $\langle{V}_{R,1} \rangle$ over the QPC gate voltage space corresponding to a transmitted single edge mode, we extract an outer edge mode transparency of $t_1> 98\%$ at the low-field end of the $\nu=2,3,4$ plateaus. For $\nu=2$, the transparency of the outer edge mode $t_1 > 99.2\%$ over the range of the quantum Hall plateau accessible with a field $< 9.5$ T. The extracted transparency of inner modes is markedly lower, likely due to their increased separation from the physical metal-semiconductor interface. Analysis and discussion of inner edge mode transparencies are included in the SM. Over each quantum Hall plateau $(\nu \in \{6, 5, 4, 3\} )$, we observe a decrease in transparency of the edge modes with increasing field. This pattern requires further investigation, but we speculate it relates to field-dependent inter-edge mode equilibration (see SM), together with the lower transparency of inner edge modes compared to the outer modes. Drive current for this measurement was $20$ nA, at a frequency below 20~Hz.

\subsection*{Static Coulomb blockade in the weak coupling limit}
\label{sec:staticCB}
\begin{figure*}[!htbp]

     \centering
         \includegraphics[width=1.0\textwidth]{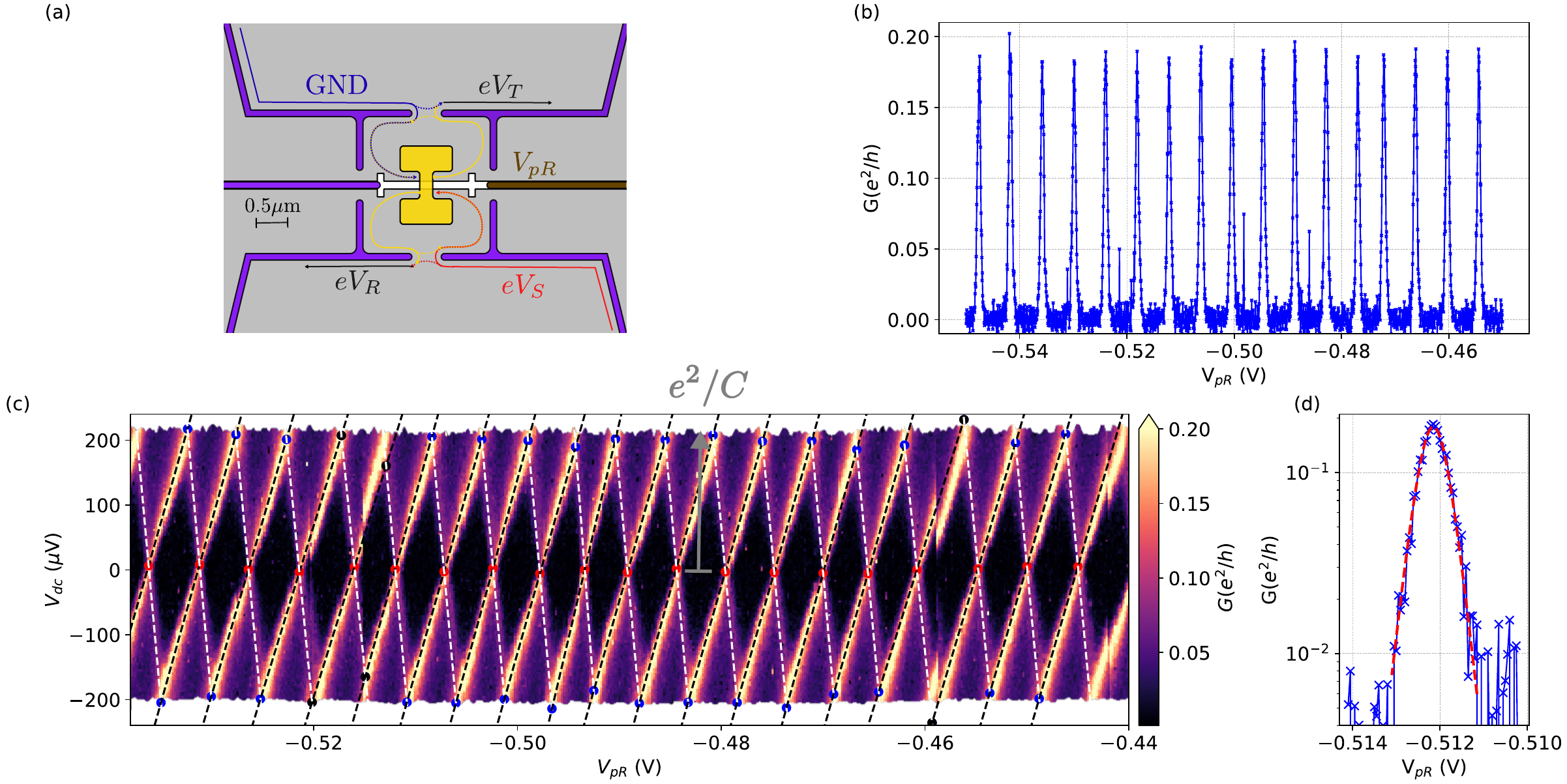}

        \caption{\textbf{Static Coulomb blockade}. (a) Schematic of the hybrid metal-semiconductor
        quantum dot (Device B). An \textit{I}-shaped Ti/Au island with a $0.8\times1.2$ $\upmu$m$^2$ footprint makes sidewall contact to the InAs quantum well across a trench(white) on a mesa (gray). The trench is 400 nm deep and 120 nm wide at its bottom, but it widens toward the top, with nearly 45$^\circ$ sidewalls. The QPC and left plunger gates (purple) are set to fixed voltages, while the right plunger (red) is swept to reveal Coulomb blockade oscillations. The device is biased with a 500 pA low-frequency ($<30$~Hz) ac excitation. A pair of contacts on either end are tied to ground at the mixing chamber stage of the dilution fridge, and realize a ``cold'' ground to drain current. Lock-in amplifiers are used to measure source ($V_S$), reflected ($V_R$), and transmitted ($V_T$) voltages $\approx$ 45-$\upmu$m from the island, relative to ground. (b) Coulomb blockade oscillations in the series conductance $G$ as the plunger gate voltage $V_{\text{pR}}$ is swept to add 17 successive electrons to the dot, one at a time. The QPCs are tuned to the weak tunneling limit $\tau_{1, 2} \ll 1$. The high density of states at the metallic island results in uniform peaks whose amplitude $G_{max} = 0.179 \pm 0.003$ $e^2/h$ depend on $\tau_{1, 2}$ and whose width is proportional to the ratio of temperature $T$ to the capacitive lever arm $\alpha$. (c) Coulomb diamonds measured by adding a swept dc current $|I_{\text{dc}}| < 25$ nA to the ac current bias. The source dc voltage drop $V_{\text{dc}} = I_{\text{dc}} \times h/3e^2$ is measured using a multimeter. The charging energy estimated from the diamond heights $E_c = e^2/2C = 100 \pm 3$ $\upmu eV$, where $C$ is the capacitance of the hybrid dot. (d) A Coulomb peak from (b) on a semi-log scale fit to Eq.~\ref{eq:Kulik} to determine the electron temperature = $54$~mK.}
        \label{fig:CB}
\end{figure*}

The device to study charge quantization (Device B) in a hybrid metal-semiconductor quantum dot consists of a similar submicron Ti/Au ohmic island flanked by QPC gates ($\approx$ 400 nm away, see Fig.~\ref{fig:CB}(a)). The hybrid dot is formed by applying a negative voltage to the QPC gates to deplete the quantum well underneath and confine electrons to small regions of InAs 2DEG around the metal island. This enables operation in a regime where thermal fluctuations of the hybrid dot's charge state are suppressed except near a Coulomb blockade peak: $k_B T_e \ll E_c = e^2/2C$, where $C$ is the total capacitance of the hybrid dot. We tune both QPCs to the weak-tunneling limit and sweep the plunger gate $V_{\text{pR}}$ to measure thermally-broadened Coulomb resonances. The electronic level spacing in the metallic island is $\varepsilon = \frac{1}{\rho_F \Omega }$, where $\rho_F$ is the electronic density of states per unit volume at the Fermi level and $\Omega$ is the volume of the island. For gold with $\rho_F \approx 10^{47} J^{-1} m^{-3}$ and $\Omega \approx 0.15$ $\upmu m^3$, the electronic level spacing $\varepsilon \approx 4\times 10^{-4}~\upmu$eV $= k_B \times 5 \times 10^{-3}$mK. Since the electron temperature is at least three orders of magnitude larger than $\varepsilon$, we can approximate the metallic island with a continuum electronic level spectrum. Transport through the hybrid dot is described by inelastic sequential tunneling through the QPCs, with the conductance across a Coulomb resonance given by Kulik and Shekhter~\cite{Kulik1975},
\begin{equation}
    G ^{\text{theory}} = \frac{G_{\infty}}{2}\frac{\alpha \delta V_{\text{pR}}/k_BT}{\sinh{\left(\alpha \delta V_{\text{pR}}/k_BT\right)}}.\label{eq:Kulik}
\end{equation}

$G_{\infty} = e^2 \rho_{\text{dot}} \left(\Gamma_1^{-1} + \Gamma_2^{-1}\right)^{-1}$ is the classical series conductance through the QPCs with tunneling rates $\Gamma_{1, 2}$ respectively, $\rho_{\text{dot}}$ is the electronic density of states at the Fermi level in the dot, $\delta V_{\text{pR}}$ is the plunger gate offset from charge degeneracy, and the plunger gate lever arm $\alpha$ is the ratio of the plunger gate capacitance to the total capacitance of the dot.
In a typical few-electron semiconductor quantum dot, $\rho_{\text{dot}}$ strongly depends on the confining potential landscape, making it very sensitive to disorder in the underlying heterostructure. For a given Coulomb blockade peak electrons tunnel to and from a particular localized electron wavefunction, with rate proportional to the electron wavefunction overlap between the dot and the corresponding lead. This overlap varies strongly across different electronic states~\cite{Beenakker1991, Fuhrer2003}, resulting in a broad distribution of Coulomb peak spacings and heights~\cite{Kouwenhoven_1998, Kouwenhoven_2001}. In our hybrid dot, the metallic island's quasicontinuum density of states results in a \textcolor{black}{nearly-}uniform peak height determined by the QPC transmissions, and \textcolor{black}{nearly-}uniform peak spacing $\Delta$ proportional to the charging energy $\Delta = 2E_c/e\alpha$. Casting the tunnel rates in terms of the measurable edge state transmissions through the QPCs, $\rho_{\text{dot}} h\Gamma_{1, 2} = \tau_{1, 2}$, the peak conductance is expressed as

\begin{equation}
    \frac{G_{\infty}}{2} = \frac{1}{2}\frac{e^2}{h}\left(\tau_1^{-1} + \tau_2^{-1}\right)^{-1},
\end{equation}

The uniformity in the Coulomb peak amplitudes over 17 successive Coulomb peaks is shown in Fig.~\ref{fig:CB}(b). The peak amplitude $G_{\text{peak}} = 0.179 \pm 0.003$ $e^2/h$, computed by fitting every peak and reporting the standard deviation weighted by the uncertainty of each fit. Equation~\ref{eq:Kulik} fits the peak for an electron temperature $T_e =54$~mK (Fig.~\ref{fig:CB}(d)). The charging energy $E_c\equiv e^2/2C$ is extracted by adding a dc bias current to the ac excitation, and measuring the Coulomb diamonds shown in Fig.~\ref{fig:CB}(c) as a function of $V_{\text{pR}}$ and the measured dc voltage drop $V_{\text{dc}}$ across the device. The maximum $V_{\text{dc}}$for which Coulomb blockade persists is given by the height of the diamonds, corresponding to the energy scale between a pair of consecutive Coulomb peaks, $e^2/C$~\cite{kouwenhoven_single_1991}. We report $E_c = 100 \pm 3$ $\upmu$eV as half the diamond height, measured by fitting straight lines through peak trajectories, with the uncertainty given by the standard deviation over 16 Coulomb diamonds. The charging energy reported here represents an approximately fourfold enhancement over hybrid dots on a GaAs/AlGaAs platform~\cite{jezouin_controlling_2016, Pouse2023}. The plunger-gate tunability of the charge, the uniformity in $E_c$ estimated from the diamond heights, and the lack of an excited electronic level spectrum are tell-tale signatures of a hybrid metal/semiconductor quantum dot. Because of the lack of a Schottky barrier at metal/n-InAs interfaces~\cite{woodall_ohmic_1981}, we are able to make the metal island much smaller (radius $\approx 0.6$ $\upmu$m compared with that in previous hybrid dot devices ($\approx 3 $ $\upmu$m in Refs.~\cite{Pouse2023, jezouin_controlling_2016}), while \textcolor{black}{still achieving} high electronic transparency at the metal/semiconductor interface. We cannot make our metal island much smaller than this because of the $400$ nm deep isotropic wet-etch we use to define a trench to be spanned by the metal (see Fig.~\ref{fig:transparency}(c)). There is room to boost $E_c$ further: more anisotropic etch profiles, and heterostructures compatible with shallower etches, are feasible, as are gate dielectrics with lower $\kappa$ than that of our ALD-grown HfO$_2$.

\subsection*{Charge Noise \label{sec:chargenoise}}
Charge quantization in the weak tunneling limit makes the conductance through the hybrid dot a function of local chemical potential. Fluctuations in local electrostatic potential -- and hence equal and opposite shifts in local chemical potential -- result from state changes in metastable two-level fluctuators (TLFs)~~\cite{Paldino2014, Kogan_1996} in the MBE-grown heterostructure and the gate dielectric. These may be based on unintentional charged donor impurities, vacancies, and/or interface charge traps~\cite{HuaIEEE,FLiefrink_1994,Ramon2010, Dekker1991}.  \textcolor{black}{In the context of quantum simulation of the Kondo lattice model, fluctuations in the local chemical potential are important because they directly modulate the charge pseudospin splitting and can limit the stability of the quantum critical point~\cite{Pierre3CK, Pouse2023}.}

\textcolor{black}{The manifestations of such ``charge noise'' in our devices include abrupt conductance jumps~\cite{MPioroLadrier, DHCobden1992, SMITH1996656}, low-frequency drift in Coulomb blockade peak locations in gate sweeps, and likely broadening of measured Coulomb blockade peaks caused by TLF switching on time scales shorter than our measurement of a single conductance data point. This last phenomenon would also distort measurements of conductance scaling near a quantum critical point that is sensitive to the local chemical potential~\cite{Diehl2008}. Charge noise spectroscopy is a valuable tool for diagnosing dominant switching mechanisms, comparing materials and processing techniques, and informing measurement protocols for noise immunity~\cite{Viola1999, biercuk_optimized_2009, Connors2019}.}

\begin{figure*}[!htbp]

     \centering
         \includegraphics[width=1.0\textwidth]{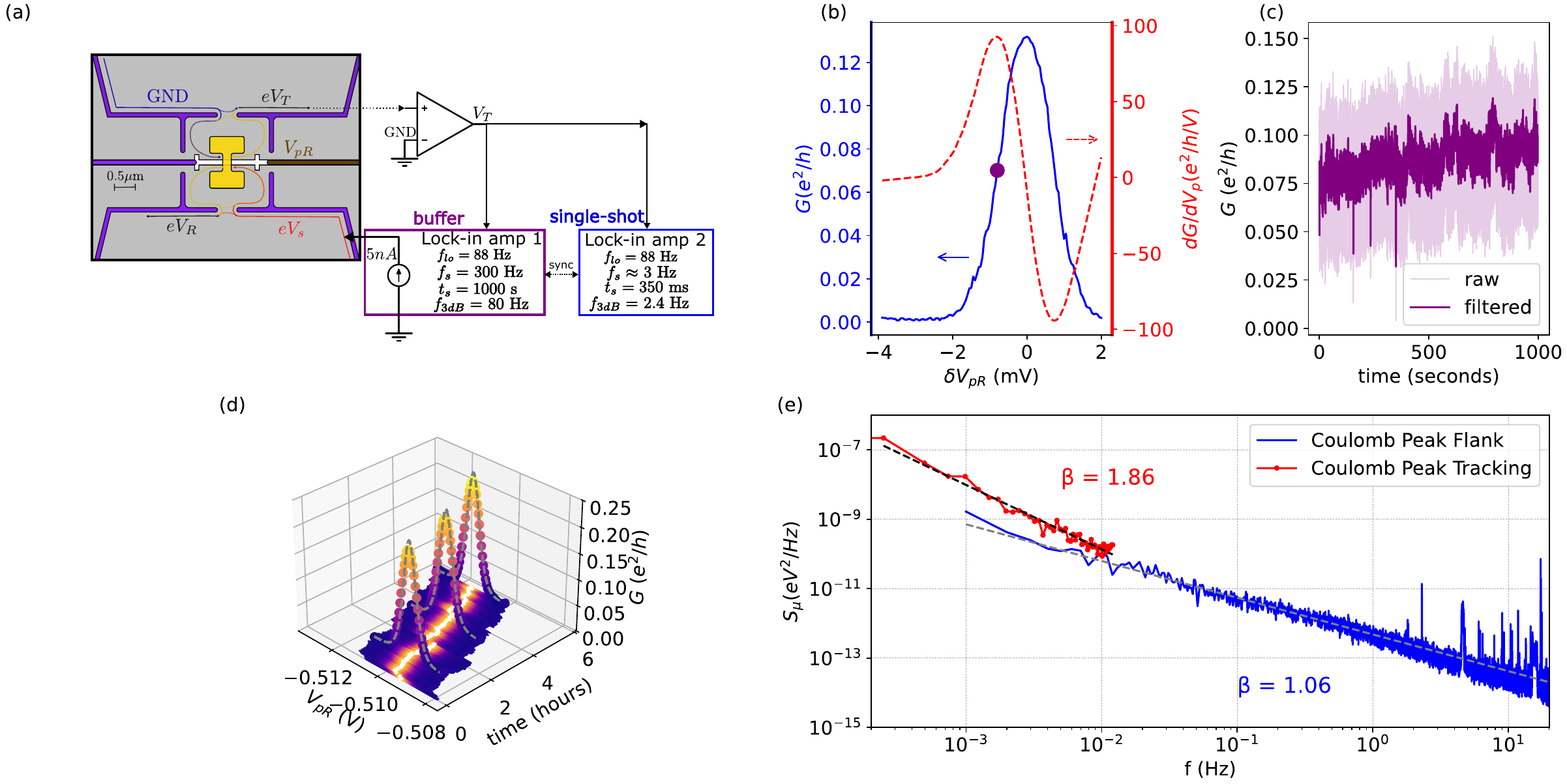}

        \caption{\textbf{Charge noise}. (a) Schematic representation of the measurement setup. The device is biased with a 5 nA low-frequency ac excitation, resulting in an on-chip voltage excitation of 5 nA$\times$ 25813/$
        \nu$ $\Omega \approx 43$ $\upmu$V for $\nu=3$. After pre-amplification, the time-series data of the voltage drop $V_T$ between the corresponding voltage probe and on-chip cold-ground is recorded in a lock-in amplifier internal buffer, for 1000 seconds, sampled at 300~Hz, with a time-constant of 3 ms. Low-bandwidth single shot measurements are recorded with a synced lock-in for resolving the Coulomb peak and identifying the flank -- the plunger gate tuning corresponding to the maximum transconductance $dG/dV_{\text{pR}}$. (b) Representative Coulomb peak (blue) and transconductance (red) as a function of plunger gate voltage $V_{\text{pR}}$. In the Coulomb peak flank method, the plunger gate is parked on the flank of the Coulomb peak (magenta circle), where the transconductance is maximized. (c) (Pink) Raw
        time series data of $G$ at the flank of the Coulomb peak. (Purple) These same data filtered to reject noise from the setup at frequencies > 20~Hz. (d) Coulomb peak tracking  measurements highlighting 3 Coulomb peaks, each fit to Eq.~\ref{eq:Kulik}. The white points on the heatmap indicate the peak positions used to calculate the low frequency PSD. (e) PSD of the charge noise extracted from Coulomb peak flank (blue) and Coulomb peak tracking (red), each fit to a separate $f^\beta$ power-law. The best-fit $(\beta = 1.06)$ for the Coulomb peak flank method resembles a $1/f$ spectrum. That for Coulomb peak tracking $(\beta=1.86)$ resembles $1/f^2$.}
        \label{fig:chargenoise}
\end{figure*}

\textcolor{black}{We investigate the charge noise spectrum over 6 decades in frequency from $100$ $\upmu$Hz to $100 $~Hz, extract the dominant scaling behavior, and estimate its impact on Coulomb peak widths and on the stability of quantum critical ground states in the charge-Kondo model. At frequencies above 10 mHz, we use the Coulomb peak flank method, in which the plunger gate voltage is parked at the steepest rise of a Coulomb peak and a time-trace of the conductance is then recorded~\cite{Connors2019, paquelet_wuetz_reducing_2023}. The measurement setup is illustrated in Fig.~\ref{fig:chargenoise}(a), with a representative conductance peak $G(V_{\text{pR}})$, and the corresponding transconductance $dG/dV_{\text{pR}}$ shown in Fig.~\ref{fig:chargenoise}(b), and corresponding time-series data in Fig.~\ref{fig:chargenoise}(c). At frequencies below 10 mHz we resort to the Coulomb peak tracking method, where fluctuations in the peak position form the time-series data~\cite{massai_impact_2024} as shown in Fig.~\ref{fig:chargenoise}(d). Both the Coulomb peak flank method and the Coulomb peak tracking method are well-established in the literature, so in the following discussion we focus on measurement results, reserving description of methodological details to Methods.}

\textcolor{black}{We observe a power law $1/f^{\beta}$-like noise spectrum (Fig.~\ref{fig:chargenoise}(e)), with exponent $\beta$ changing at $\approx 10^{-2}$Hz. Above this frequency, the exponent $\beta=1.06\approx1$ is a signature of an ensemble of TLFs with switching rates nearly-uniformly distributed on a log scale. This could be explained by thermally-activated TLFs with nearly-uniformly-distributed activation energies ~\cite{Kingston2012, DuttaHorn}. Below this frequency, the exponent $\beta = 1.86\approx2$ is a signature of charge noise dominated by only one or perhaps a few TLFs~\cite{Vandersypen2008, Connors2019}.} 

\textcolor{black}{Our measured charge noise of $0.45\pm0.11~\upmu\text{eV}^2/\text{Hz}$ at 1~Hz, averaged over 14 peaks, compares favorably to the lowest reported value of $1~\upmu\text{eV}^2/\text{Hz}$ at 1~Hz for quantum dots in III-V materials~\cite{basset_evaluating_2014, jekat_exfoliated_2020}. For our many-electron hybrid dot, we observe no strong dependence of charge noise on the choice of Coulomb peak; see SM for individual peak PSDs. This differs from reports on charge noise in few-electron semiconductor dots~\cite{paquelet_wuetz_reducing_2023}. The TLFs are likely associated with unintentional dopants in the heterostructure~\cite{Hatke2017} and/or charge traps in the 35 nm ALD-grown HfO$_2$ dielectric. The low end of the $0.04$–$0.88~\upmu\text{eV}^2/\text{Hz}$ range of charge noise reported for single-electron quantum dots in undoped Si/SiGe heterostructures~\cite{XMi2018, connors_charge-noise_2022,stehouwerExploitingStrainedEpitaxial2025} is even lower than we find, possibly because of fewer unintentional dopants and higher-quality dielectrics and interfaces.}

\textcolor{black}{Though the charge noise at 1~Hz is useful as a standard benchmark, the device is affected by the total noise over the bandwidth corresponding to the time scales of a measurement. From our measured noise spectrum, we determine the total RMS chemical potential noise over a bandwidth of interest}

\begin{equation}
\sigma_{\upmu} = \sqrt{\int_{f_{0}}^{f_{1}} S_{\upmu}(f) df}.
%= 2.2\ \upmu\text{eV},
\label{eq:rmsmunoise}
\end{equation}

\textcolor{black}{For spin qubits, the most relevant bandwidth is the range over which noise can decohere the qubit state~\cite{KDPetersson2010}. In our case, we're more concerned by fluctuations in the chemical potential broadening or distorting observed Coulomb blockade peaks. The lower cutoff frequency for noise to cause such broadening is set by the inverse of the time taken to measure conductance at a single gate voltage, $f_0 = 2.86$~Hz. The upper cutoff is set by the highest tunneling rate between dot and reservoir with QPC gates at a particular setting. A conservative upper bound for such tunneling rates would be the full width at half maximum of our Coulomb blockade peak $f_1 = k_B T_e/h \approx 1$ GHz -- that is likely an overestimate, since peaks in this regime appear to be thermally-broadened rather than lifetime-broadened. Over this bandwidth, we find $\sigma_{\upmu} = (2.2 \pm 0.3)$ $\upmu\text{eV}$.}

\textcolor{black}{Though fluctuations in the chemical potential can lead to broadening of the Coulomb peak lineshape, the estimate from Eq.\ref{eq:rmsmunoise} corresponds to only a 1.5\% increase in the full width at half maximum (FWHM) relative to the ideal thermal lineshape given in Eq.\ref{eq:Kulik} for $T_e=54$~mK (see SM). This indicates that thermal broadening remains the dominant broadening mechanism, supporting the use of Coulomb peak width as a proxy for electron temperature.}

\textcolor{black}{Finally, we quantify the implications of our estimated $\sigma_{\upmu}$ for the stability of fragile quantum critical points in the context of quantum simulation. The ratio $\sigma_{\upmu}^2 / E_c = 48~\text{neV} \approx  0.6$~mK$k_B$ defines a characteristic temperature scale $T^*$, below which charge noise may become relevant for the stability of the critical point in the charge Kondo model~\cite{Mitchell2016}. A useful criterion for acceptable charge noise is $T^*\ll$ the electron temperature $T_e$, so that noise-induced charge fluctuations remain small compared to thermal broadening and do not influence quantum critical behavior. Since $T^*$ is more than an order of magnitude smaller than the minimum electron temperature achieved in our measurement ($T_e = 54$~mK), we expect the quantum critical point to remain robust against charge noise in future investigations of lattice models based on charge Kondo interactions. However, were $T_e$ reduced to state-of-the-art levels $\approx10$~mK the constraint on acceptable charge noise would become correspondingly more stringent.}

\subsection*{Charge quantization in the hybrid dot in the nearly-ballistic limit}
\label{sec:chargeQ}
\textcolor{black}{In the previous section, the tunnel coupling of the hybrid dot to its leads was set to be weak, resulting in a discrete many-body spectrum characterized by the charge of the hybrid dot~\cite{Beenakker1991} and suppressing charge fluctuations at temperatures $k_BT_e \ll E_c$. Stronger coupling to the leads allows the charge on the hybrid dot to fluctuate~\cite{kouwenhoven_single_1991, Amasha2011, Berman1999, Chouvaev1999, Joyez1997, STARING1991226}. The ability to smoothly tune edge mode transmissions $\tau_{1,2}$ through the QPCs from the weak to strong tunneling limit is a key requirement for quantum simulation on this platform.}

\textcolor{black}{As depicted in Fig.~\ref{fig:CB}(b), charge quantization results in conductance oscillations as a function of $V_{\text{pR}}$. The visibility of conductance oscillations is given by}

\begin{equation}
    Q = \frac{\max G(V_{\text{pR}})- \min G(V_{\text{pR}})}{\max G(V_{\text{pR}}) + \min G(V_{\text{pR}})},
\end{equation}
\textcolor{black}{where $\max$ $(\min)$ are the maximum (minimum) conductance over one $V_{\text{pR}}$ period $\Delta$. This visibility $Q$ is directly proportional, with a coefficient of order unity, to the visibility of charge oscillations accessible in capacitance measurements~\cite{YiKane1996, jezouin_controlling_2016}.}

\begin{figure*}[!htbp]
     \centering
         \includegraphics[width=1.0\textwidth]{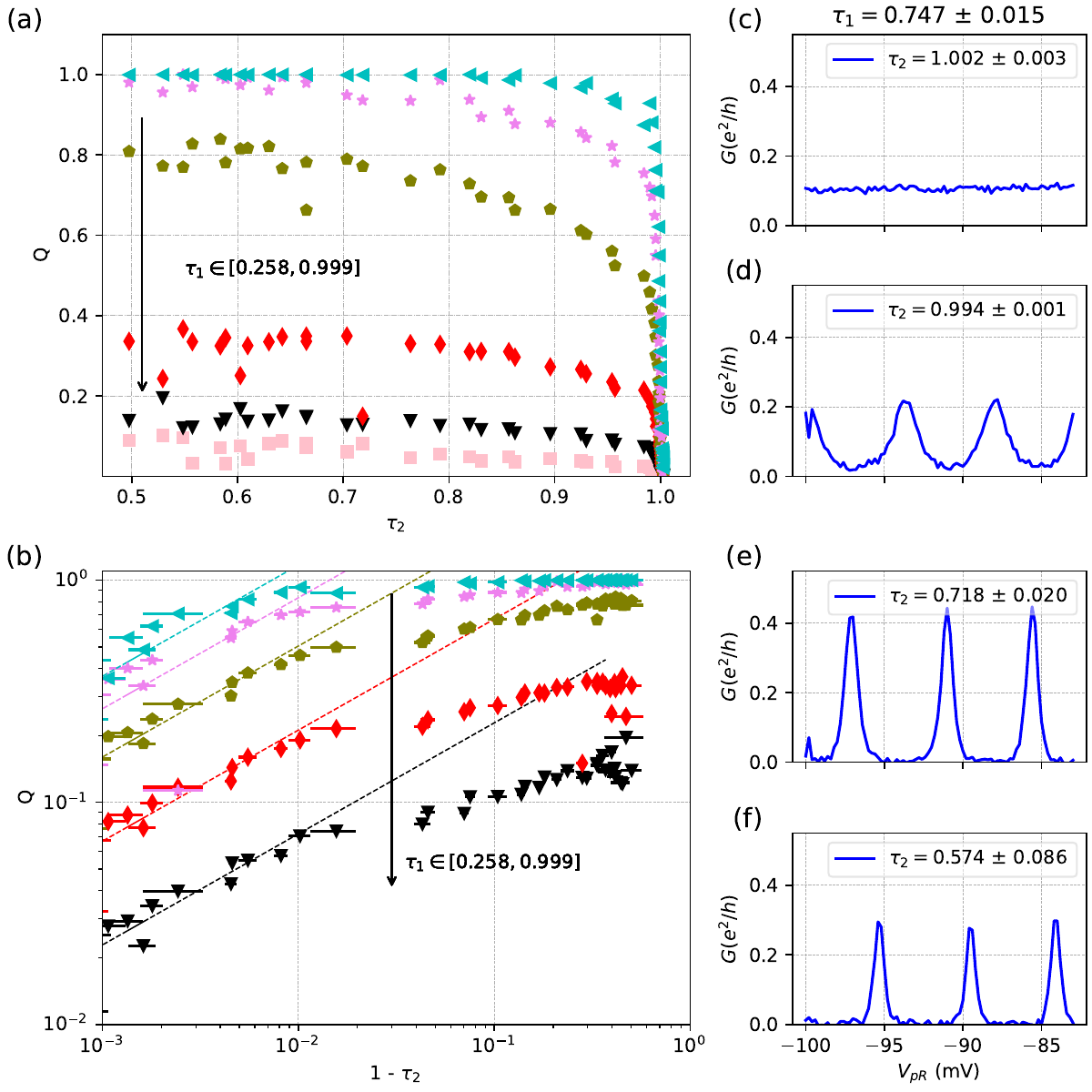}

        \caption{\textbf{Controlling charge quantization} (a) The visibility of conductance oscillations $Q$ as a function of top QPC transmission $\tau_2$, for several fixed bottom QPC transmissions $\tau_1 \in$  $[0.258, 0.999]$. (b) The evolution of $Q$ as $\tau_2$ is tuned away from the ballistic limit $(1-\tau_2 \ll 1)$. Dashed lines: The asymptotic $\sqrt{1 - \tau_2}$ scaling, with the prefactor (vertical offset of each line on this log-log plot) left as a fit parameter. See Methods for details. (c-f) Series conductance through the hybrid dot for fixed $\tau_1 = 0.747 \pm 0.015$, as $\tau_2$ is varied $\in \{1.002, 0.999, 0.718, 0.574\}$ between the panels, with uncertainties as indicated.}
        \label{fig:chargeq}
\end{figure*}

\textcolor{black}{Representative conductance oscillations $G(V_{\text{pR}})$ as a function of $\tau_2$ for a fixed $\tau_1 = 0.747 \pm 0.015$ are shown in Fig.~\ref{fig:chargeq}(c)-(f). For each panel, $\tau_2 $ is set to a different value by adjusting the top QPC gate voltage. The protocol for determining the values of $\tau_{1,2}$ is described in the Methods, and the corresponding dataset is shown in the SM.}

\textcolor{black}{In the ballistic limit ($\tau_2=1$), fluctuations in charge on the hybrid dot are of order the electron charge. Charge quantization is quenched, as seen by the fact that the measured $G$ is independent of $V_{\text{pR}}$ (see Fig.~\ref{fig:chargeq}(c)), i.e. $Q=0$. 
% In the ballistic $(\tau = 1)$ limit of either QPC, quantum fluctuations entirely quench this quantization~\cite{jezouin_controlling_2016}. 
This is in contrast to semiconductor dots, where the interplay of coherence and Coulomb interactions preserves weak ``mesoscopic Coulomb blockade'' conductance oscillations periodic in the average dot charge even in the limit of a fully transmitting mode through each QPC~\cite{Amasha2011}. The continuum single-particle electronic level spectrum of the hybrid dot $(\varepsilon \ll k_BT_e)$ suppresses elastic cotunneling across the Coulomb barrier~\cite{AVERIN1989251,Nazarov1990}, so each QPC should independently control one intersite tunnel coupling for quantum simulation in a lattice geometry.}  

\textcolor{black}{As the top QPC is tuned away from the ballistic limit $(\tau_2 < 1)$, charge fluctuations are suppressed and $G(V_{\text{pR}})$ oscillates with the period $\Delta$ characteristic of Coulomb blockade. Coulomb peaks separated by valleys of near-zero conductance are a signature of robust charge quantization, seen over a wide range of transmissions $\tau_2\leq0.90$, as also reported by Jezouin et al.~\cite{jezouin_controlling_2016} for hybrid dots built on a GaAs/AlGaAs heterostructure.}~\textcolor{black}{Figure~\ref{fig:chargeq}(a) shows the evolution of charge quantization as $\tau_2$ is smoothly tuned away from the ballistic limit $(\tau_2 = 1)$ to $\tau_2 \gtrsim 0.5$. Each symbol represents a different value of $\tau_1 \in [0.258, 0.999]$. In the weak backscattering limit $(1-\tau_2 \ll 1)$, charge fluctuations are dominated by tunnel coupling through the top QPC, resulting in a strong dependence of $Q$ on $\tau_2$. As the transmission through the top QPC is further reduced $(\tau_2 \lesssim 0.7)$, charge fluctuations through the top QPC become strongly suppressed, and $Q$ becomes only weakly dependent on $\tau_2$. In this regime, the visibility is instead primarily controlled by the transmission through the bottom QPC, $\tau_1$. These measurements demonstrate continuous electrostatic control of the tunnel coupling between the hybrid dot and leads, with access to both the weak and strong coupling limits within a single device.}

\textcolor{black}{Returning to the weak backscattering limit for one lead $(1-\tau_2 \ll 1)$, a bosonization approach predicts that charge quantization should scale as $\sqrt{1-\tau_2}$ for $k_BT_e \ll E_c$, irrespective of the transmission $\tau_1$ of the other lead~\cite{Flensberg1993}. To highlight this scaling, Fig.~\ref{fig:chargeq}(b) plots $Q$ as a function of $1-\tau_2$ on a log-log scale. The expected asymptotic $\sqrt{1 - \tau_{2}}$ scaling is shown as dashed lines with slope 1/2 on this log-log plot. The qualitative match of these lines to the data for $1-\tau_2 \ll 1$ confirms that this scaling holds in the nearly ballistic limit. Deviations from predictions are likely due to renormalization by dynamical Coulomb blockade, as discussed in the SM. Though the dependence of the vertical offset on $\tau_1$ is in principle known, we have left this offset as a free parameter as we fit to the data for each bottom gate voltage setpoint, to account for drift in its mapping to $\tau_1$.} 

\subsection*{Quantum Simulation\label{sec:simulation}}

Building on our demonstration of a single hybrid metal-InAs quantum dot with a highly transparent metal contact, smoothly tunable tunnel couplings to reservoirs, and a charging energy that is high compared to the thermal energy, \akm{we now discuss how arrays of coupled hybrid dots can be used as a platform for quantum simulation. We focus here on simulating 1D models of strongly correlated quantum matter, implemented in a linear array of hybrid dots.} 

In the schematic depiction of Fig.~\ref{fig:ssh_kondo_device} each site in the linear array is confined by gate electrodes, which deplete the InAs 2DEG below. Each site also incorporates a metallic island with highly-transparent interfaces to the neighboring 2DEG, forming a hybrid dot with quasi-continuum density of states. A plunger gate on each hybrid dot can be tuned to set the local charge to a degeneracy point. QPCs tune inter-dot tunnel couplings by controlling quantum Hall edge mode transmissions. As noted above, and demonstrated in our single-hybrid-dot device, this architecture facilitates smooth and independent control of each separate tunnel coupling.

Just as with our single-hybrid-dot demonstration, we envision operating at high magnetic field, with the InAs 2DEG in the quantum Hall regime. Chirality of quantum Hall edge modes will suppress backscattering within the semiconductor regions of each dot, so that the effective tunnel coupling between two neighboring dots is determined by the transmission of a quantum Hall edge mode through the inter-dot QPC. Each such inter-dot coupling is expected to be tunable and equal in both directions, as the edge mode can travel through the QPC both from left to right and back from right to left. Thus, chirality provides a practical advantage without constraining the geometry of lattices that can be realized in this architecture. Such symmetric and tunable inter-dot coupling through a chiral edge mode has already been demonstrated by Pouse {\em et al.}~\cite{Pouse2023} in a double hybrid dot based on a GaAs 2DEG.

\textcolor{black}{An important distinction from an array of conventional semiconductor quantum dots~\cite{Simmons2022} is that each site in the hybrid dot array brings with it a screening reservoir associated with the metal island's continuum single-electron spectrum~\cite{Matveev1991QuantumReservoir}. Thus, as demonstrated in the double hybrid dot system by Pouse {\em et al.}~\cite{Pouse2023}, neighboring pairs of dots are coupled not through a simple antiferromagnetic coupling but rather through a Kondo interaction. Arrays of such sites map onto a family of Kondo-lattice Hamiltonians~\cite{Pouse2023}, relevant to understanding materials including heavy fermion compounds.}

\eran{Below we describe some of the range of models that can be simulated using such an array.  See SM for a more in-depth description of the Hamiltonian for such an array. A 1D array exhibits a variety of regimes depending on the strength of transmission between neighboring dots. Perfect QPC transmission ($\tau=1$) from each dot to its neighbor along the chain is expected to yield a metamaterial with nonlocal response properties~\cite{Maillet2026,Sela2026}. Turning on partial backscattering ($\tau<1$) at a single point along the chain is predicted to produce fractional charge tunneling with an effective Luttinger liquid parameter controlled by the total number of dots in the chain~\cite{Karki_2022}. Such tunneling has already been studied experimentally in the context of a single hybrid dot~\cite{Anthore_2018}. Turning on finite and equal backscattering at two points along the chain is expected to realize a generalization of the Majorana fermions that appear in the two-channel Kondo model: parafermions in the two-dot case~\cite{Karki2023}, and more general anyonic excitations with fractionalized statistics for arrays with more than two dots~\cite{Sela2026}.} \akm{The richest and most complex physics is expected when interactions are strong and backscattering is finite at not one or two points along the chain, but at all points. This is precisely the regime where classical computation fails, and experimental quantum simulators can provide unique insights. In the next section we discuss a specific example of this, and the nontrivial physical regimes that we expect a realistic simulator device should be able to access.}

\subsection*{Simulator Device Proposal}\label{sec:proposal}

\begin{figure*}[!htbp]%[t] % or [h], [b], etc.
    \centering
    \includegraphics[width=0.65\textwidth]{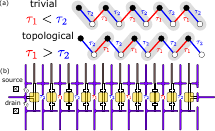}
    \caption[Hybrid dot linear-array device for realizing an interacting variant of the topological SSH model]%the SSH-Kondo model]
    {\textbf{Hybrid dot linear-array device for realizing an interacting variant of the topological SSH model.} % the SSH-Kondo model.} 
(a) Finite-length one-dimensional hybrid quantum dot chain with alternating couplings $\tau_1$ and $\tau_2$. The chain is expected to exhibit a trivial phase for $\tau_1 < \tau_2$ (top) and a \textit{topological} phase for $\tau_1 > \tau_2$ (bottom). 
(b) Realistic schematic gate layout of such a chain. Purple QPC gates tune interdot transmissions $\tau_1$ and $\tau_2$, while brown plunger gates tune the local chemical potential to reach charge degeneracy at each site. The chain and its topological states can be probed by tunneling between a pair of leads, both coupled to the same site at one end of the chain, with coupling $\tau_0\ll\tau_1,\tau_2$.}
    \label{fig:ssh_kondo_device}
\end{figure*}

\akm{The Su-Schrieffer-Heeger (SSH) model~\cite{SSH1979} is a famous paradigm for non-interacting topological quantum matter in 1D. Here we argue that the site-to-site tunability of a 1D array of coupled hybrid dots allows for a nontrivial \textit{interacting} variant of this model to be simulated experimentally.

The standard SSH model consists of a finite 1D chain of fermionic sites, with tunnel-couplings alternating between two values $t_1$ and $t_2$ along the chain. Strongly-coupled neighbors form a series of dimers along the chain, with one dimer weakly coupled to the next. If the site at each end of the chain is weakly coupled to its neighbor, these end sites become free, effectively decoupled from the rest of the system. This is the `topological' phase, in which  boundary-localized zero-mode states are predicted to be topologically-protected. By contrast, in the `trivial' phase where the end sites are incorporated into dimers there are no such boundary states. The situation is more subtle when electron interactions are introduced~\cite{intSSH} but in the strongly-correlated limit of a Hubbard-SSH model an effective chain of spins with alternating antiferromagnetic exchange couplings is expected, with free spin-$\tfrac{1}{2}$ local moments at the end of the chain in the topological phase~\cite{zhao2024tunable}.

In the proposed experimental quantum simulator depicted in Fig.~\ref{fig:ssh_kondo_device}, the coupling strength between neighboring hybrid dots in a 10-site linear array is controlled by the transmission from one to the other through the intervening QPC, and this transmission alternates between two values, $\tau_1$ and $\tau_2$, along the chain. Plunger gates tune the local
chemical potential of each dot to a charge degeneracy point, which implements a charge pseudospin consisting of $N$ and $N+1$ electron charge states. 
Electron tunneling between connected dots therefore produces a mutual pseudospin flip on these dots, since the charge of one dot increases by one unit as the charge of the other correspondingly decreases. However, unlike the original free-fermion SSH setup or the pure 1D Heisenberg model, here a kind of Kondo lattice chain is realized, since inter-dot couplings are of \textit{Kondo} type, rather than simple exchange, with spin-flips correlated to electronic scattering. The resulting SSH-Kondo model which our proposed device simulates should feature an interplay between local moments and itinerant electrons reminiscent of that seen in heavy fermion compounds, in which Kondo and RKKY-type exchange compete~\cite{coleman2006heavy}.

This SSH-Kondo model has yet to be explored theoretically. However, we expect a nontrivial interplay between strong correlation physics and topology. In particular, the SSH dimers now consist of Kondo-locked dots on neighboring sites, whose renormalized coupling grows as the temperature is lowered~\cite{Pouse2023}. Thus interactions should reinforce rather than reduce topological effects, driving the system deep into either a topological or a trivial phase, even when the difference between the transmissions $\tau_1$ and $\tau_2$ is small. A clear experimental signature of the boundary-localized free pseudospin in the topological phase would be a strong Kondo-boosted conductance between source and drain leads weakly attached to the same dot at one end of the chain, with transmission $\tau_0 \ll \tau_1, \tau_2$, as shown in Fig.~\ref{fig:ssh_kondo_device}(b). In the trivial phase the terminal dot is locked in a Kondo-dimer state with its neighbor, so transport through it is Kondo-blocked. Series transport through the entire chain should only be possible at the gap-closing critical point between topological and trivial phases. These expectations, and the details of the phase diagram, can be tested in our proposed simulator device.}

\section*{Discussion}
We have described the fabrication and characterization of a gate-confined hybrid metal-InAs quantum dot coupled to macroscopic reservoirs through QPCs. In the weak tunneling limit, the hybrid dot displays single-electron transport with remarkable uniformity of Coulomb peak heights. When using such hybrid dots to study quantum criticality~\cite{Mitchell2016}, a key figure of merit is the ratio of energy scales $E_c/k_BT$. Previous few-micron hybrid dot realizations based on an GaAs quantum well reached $E_c/k_BT_e \approx 26$~\cite{Iftikhar2015} and $\approx 15$~\cite{Pouse2023} \textcolor{black}{in carefully-filtered dilution refrigerators with $T_e \approx 10-20$~mK.} In the present work we instead use an InAs quantum well, enabling near-perfect transparency of the outermost quantum Hall edge mode to even a much smaller, submicron metallic island. The $100\text~\upmu$eV charging energy extracted from the Coulomb diamonds is substantially larger than in GaAs-based hybrid dots, allowing us to reach a comparable $E_c/k_BT_e \approx 22$ even at our much higher base electron temperature $T_e= 54$~mK of our cryostat. \textcolor{black}{Reaching a comparable operating regime with less demanding cryostat requirements and with a nearly 20X reduction in footprint compared to previous implementations~\cite{Iftikhar2015,Pierre3CK,Pouse2023} in GaAs-based platforms is attractive for scaling up toward lattice arrays for quantum simulation.} Scaling down the size of the hybrid dots to boost the charging energy by a further 2-3X seems feasible, using shallower and/or more anisotropic etches. And of course this $E_c/k_BT_e$ could also be boosted by measuring this style of dot in a more carefully-filtered cryostat.  Charge noise spectroscopy quantifies the effect of TLFs coupled to the dot, revealing a $S_\upmu = 0.45$ $\upmu$eV$^2$/Hz noise at 1~Hz, competitive with reported values for conventional quantum dots in various semiconductor heterostructures. \textcolor{black}{This work demonstrates uniformity of Coulomb peaks across many successive electron additions within a single hybrid dot; demonstrating such uniformity across multiple dots in an array remains an important direction for future work.}

The smooth tunability of outer edge mode transmissions through the QPCs enables a systematic study of charge quantization, demonstrating well-defined conductance peaks and valleys for $\tau_{1,2} \leq 0.7$. At the ballistic limit of either QPC, charge quantization is quenched, and the conductance is independent of the plunger gate voltage.
% although effects of Coulomb interactions persist in the form of dynamical Coulomb blockade, resulting in the low-bias $|eV_{\text{dc}}| \ll E_c$ suppression of transmission.  
This study supports the prospect of using hybrid metal-InAs quantum dots \textcolor{black}{toward simulating} the physics of multi-site Kondo models based on the charge pseudospin of the hybrid quantum dot ~\cite{FurusakiMatveev1995}. In the longer term, this approach \textcolor{black}{may provide a route toward} using such models~\cite{DONIACH1977231} to gain insights into lattice coherence observed in heavy fermion materials ~\cite{Coleman2015}.

\section*{Methods}
% \renewcommand{\thefigure}{M\arabic{figure}}
% \renewcommand{\theequation}{M\arabic{equation}}

%%%%%%%%%% Prefix a "S" to all equations, figures, tables and reset the counter %%%%%%%%%%
% \setcounter{figure}{0}
% \setcounter{equation}{0}

\subsection*{Indium Arsenide quantum well heterostructure stack}
\begin{figure*}[!htbp]%[!htbp] 
        %\begin{minipage}[b]{0.45\linewidth}
            \centering
            \includegraphics[width=\textwidth,keepaspectratio]{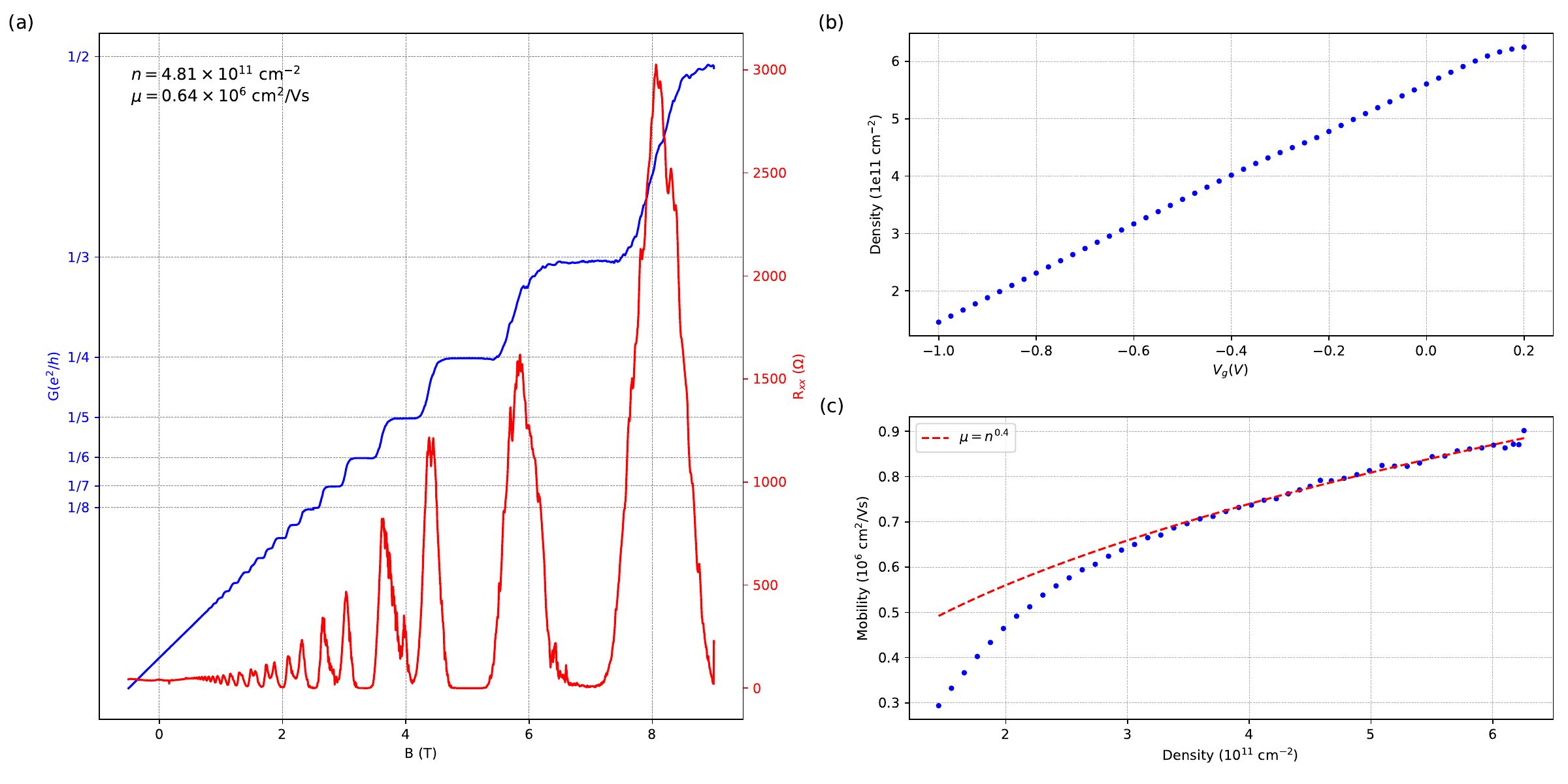}
\caption{\textbf{Transport characterization of the InAs Heterostructure.} (a) Four-terminal magnetotransport data for a $60$ $\upmu$m wide ungated Hall-bar. Extracting the density from low-field magnetoresistance in addition to Shubnikov-de Haas (SdH) oscillations yields a consistent density of 4.8$\times10^{11}$~$\text{cm}^{-2}$, confirming single-subband occupation. (b) Density modulation on a different Hall-bar device with a Ti/Au gate biased at $V_g$. Sweeping the gate results in a reduced density of carriers, and the linear dependence of density on voltage is expected from a simple capacitance model. (c) Mobility variation with density, with a $\mu \propto n^{0.4}$ power law fit to the data, suggesting unintentional background impurities and native charge defects as the main mechanisms limiting the mobility.  See text for discussion.} 
 	 \label{fig:stack}
        %\end{minipage}
      % \hspace{0.5cm}
        %\pause
            \end{figure*}

The devices featured in this work were fabricated on a heterostructure grown by molecular beam epitaxy on a semi-insulating InP (100) substrate. 
A cross-sectional schematic of the InAs quantum well (QW) layer structure is shown in the supplementary material of \cite{Hsueh2022}, with the growth parameters detailed in Ref.~\cite{Hatke2017} (Sample B). The 4 nm InAs QW sandwiched between 10.5 nm In$_{0.75}$Ga$_{0.25}$As barriers forms  the active region of the stack and hosts the two-dimensional electron gas (2DEG). The InAs quantum well is populated by unintentional impurities from deep-level donor states in the In$_{0.75}$Al$_{0.25}$As layers~\cite{Luo1993,Capotondi2004}. Simulations discussed in the supplement of \cite{Hsueh2022} built on previous work by \cite{PhysRevB.97.085302} reveal electron density concentrated in the InAs QW, with a single subband occupied. The 2DEG electron density extracted from low-field Hall resistance and Shubnikov-de Haas (SdH) oscillations in a 60 $\upmu$m wide ungated Hall-bar fabricated on the same chip corroborate the occupation of a single subband, with density $n_{\text{Hall}} \simeq n_{\text{SdH}} =  4.81\times 10^{11}$ cm$^{-2}$ and a mobility of $\mu = 6.4\times 10^{5}$ cm$^2$/Vs (see Fig.~\ref{fig:stack}(a)). This mobility is enhanced by aspects of the heterostructure design: a 900 nm linearly graded In$_{x}$Al$_{1-x}$As strain-relaxed buffer layer, and a 120 nm In$_{0.75}$Al$_{0.25}$As capping layer.

The mobility variation with density is probed on a nominally identical Hall-bar with a Ti/Au top gate to tune the density. The linear dependence of the density on the top gate voltage, expected from a simple capacitive model, is depicted in Fig.~\ref{fig:stack}(b), with the mobility dependence on density shown in Fig.~\ref{fig:stack}(c). A fit to a $\mu \propto n^{\alpha}$ power law in the Thomas-Fermi screening limit yields $\alpha = 0.4$, consistent with mobility limited by unintentional background impurities and native charged point defects~\cite{SDS_scaling2013}.%\citeSMsupp{ShabaniMIT,DasSarma_2013}.

\subsection*{Additional fabrication details}
Electron-beam lithography was performed at the Stanford Nano Shared Facilities with a 50 kV Raith Voyager using 495k PMMA A5 resist and developed in 1:3 methyl isobutyl ketone (MIBK): isopropanol. Resist removal was aided by a remote oxygen plasma descumming step. The wet mesa etchant comprised 12.01 g of citric acid anhydrous in solution with 250 mL H\textsubscript{2}O, 3 mL H\textsubscript{3}PO\textsubscript{4} (85\%), and 3 mL H\textsubscript{2}O\textsubscript{2} (30\%). The etch depth of 400 nm was chosen to avoid parallel conduction in the buffer layer which had been observed in similar heterostructures but not necessarily this one. In the ohmic contact step, following the mask development the sample was dipped in a dilute (1:9) solution of HCl to remove native oxide on the exposed InAs sidewalls. The 35-nm HfO\textsubscript{2} dielectric layer was grown by thermal ALD at 150$^\circ$C using a Cambridge Nanotech (now Veeco) Savannah.  

% \section{Device design and measurement details}
\subsection*{Instrumentation and Measurement Details}
The chip was mounted on a Quantum Machines QBoard-I, aluminum-wirebonded, and cooled down with an insertable probe in a Leiden Cryogenics CF-CS81 dry dilution refrigerator. The base temperature near the mixing chamber contact on the insertable probe was 50~mK. Current biased transport measurement were performed as shown in the schematic of the measurement setup in Fig.~\ref{fig:transparency}(a) of the main text. A low frequency ($<100$~Hz) AC 50 mV--1 V RMS sinusoidal signal was sourced from a Stanford Research Systems SR860 lock-in amplifier with a 100 M$\Omega$ bias resistor. The variable DC bias current $I_{\text{bias}}$ was provided by a Quantum Machines QDAC voltage source with another 100 M$\Omega$ bias resistor. The AC and DC bias currents were summed connected to the source terminal of the devices, and drained at both ends of the island through grounded contacts that were tied directly to the cold finger of the dilution refrigerator. The voltage drop between voltage probes downstream of the island and the grounded current drain was amplified by a factor of 100 using an NF SA-240F5 low noise FET preamplifier. The AC output of the preamplifier, $V_{\text{S}}$, $V_{\text{R}}$, $V_{\text{T}}$, and $V_{\text{I}}$, were measured using Stanford Research Systems SR830 lock-in amplifiers, and the DC outputs were measured using Keysight 34465A digital multimeters. The QPC and plunger gate voltages were sourced from a Quantum Machines QDAC-I.

\subsection*{Charge Noise Spectroscopy}

\textcolor{black}{Noise in our measured current and hence in our inferred conductance predominantly reflects fluctuations in the chemical potential, rather than noise in source-drain bias. Current noise could also be produced by noise in the source-drain bias voltage, but this would be proportional both to the noise in the bias voltage and to the conductance. As shown in SM, our measured current noise on the Coulomb flank exceeds that on the peak, even though conductance is higher on the peak.} 
% The measured charge noise resembles a $1/f^{\beta}$ spectrum with $\beta = 1.06$, and $S_{\mu}(1 \text{~Hz}) \sim 0.45 \pm 0.11$ $\upmu$e$V^2$/Hz. 

\textcolor{black}{At frequencies above 10 mHz, we use the Coulomb peak flank method, in which the plunger gate voltage is parked at the steepest rise of a Coulomb peak and a time-trace of the conductance is then recorded. The conductance noise power spectral density (PSD) $S_{G}$ is then translated to charge noise PSD $S_{\mu}$ using a linear local approximation to the Coulomb peak shape}

\begin{equation}
    S_G = \left(\alpha^{-1}\frac{\partial G}{\partial V_{\text{pR}}}\right)^2 S_{\mu},
    \label{eq:smu}
\end{equation}

\textcolor{black}{where $\alpha$ is the lever arm extracted from the Coulomb diamond measurement in Fig.~\ref{fig:CB}(c). This method assumes that excursions in potential are small enough for the linear approximation to hold~\cite{Beenakker1991}. Low-frequency drift thus limits the maximum sampling time.
Figure~\ref{fig:chargenoise}(b) shows a representative conductance peak $G(V_{\text{pR}})$, and the corresponding transconductance $dG/dV_{\text{pR}}$. $V_{\text{pR}}$ is tuned to maximize $dG/dV_{\text{pR}}$ (magenta circle), and then the $V_T$ time-trace is recorded to a buffer on the lock-in amplifier for 1000 seconds, at a sampling rate of 300 samples per second. The lock-in excitation frequency $f_{lo} = 88$~Hz, and 3-dB bandwidth $f_{3dB} = 80$~Hz were chosen to facilitate distinguishing device charge noise from noise in the measurement setup.
Figure~\ref{fig:chargenoise}(c) shows a representative raw time trace (pink), and a version filtered to reject setup noise above 20~Hz.}

\textcolor{black}{At frequencies below 10 mHz, the slow drift renders the linear approximation invalid. Consequently, we resort to the Coulomb peak tracking method, where a plunger gate voltage is repeatedly swept back-and-forth over a single Coulomb peak, and the peak position forms the time-series data~\cite{massai_impact_2024}. We record data for 10,000 seconds with sweeps that last $t_\text{meas}=47$ seconds in each direction, as shown in Fig.~\ref{fig:chargenoise}(d). Using the lever arm $\alpha$, the corresponding PSD is translated to charge noise, over a bandwidth spanning 100 $\upmu$Hz - 10 mHz. The noise PSD has power law exponent $\beta = 1.86$, resembling a Lorentzian spectrum. This suggests that only one or a few TLFs dominate charge noise in this frequency range~\cite{MPioroLadrier, DHCobden1992, SMITH1996656}.}

\subsection*{Charge Quantization in the Nearly Ballistic Limit}
\textcolor{black}{The reported $\tau_{1,2}$ values and the corresponding error bars are determined from the bias dependence of the transmissions extracted from a series conductance measurement. For the top QPC, the zero bias transmission $\tilde{\tau}_{2}(V_{\text{dc}}=0)$ is first extracted with the bottom QPC set to transmit one edge mode $\tau_1 = 1$, $\tilde{\tau}_{2}(V_{\text{dc}}=0) = h/e^2\left(G^{-1}(V_{\text{pR}}) - h/e^2\right)^{-1}$, while sweeping the plunger gate over a range corresponding to three Coulomb peaks, as shown in Fig.~\ref{fig:chargeq}(c). The error bars in $\tilde{\tau}_{2}(V_{\text{dc}}=0)$ are determined from the standard deviation of the series conductance over the swept plunger gate range. The low-energy intrinsic transmission $\tau_2$ is subsequently predicted from $\tilde{\tau}_{2}(V_{\text{dc}}=0)$ by fitting the low-bias $\tilde{\tau}_{2}(V_{\text{dc}})$ to Eq.~\ref{eq:DCBmain}, as described in SM, with the corresponding data presented in SM Fig.~\ref{fig:DCB_B}. The SM section on Dynamical Coulomb Blockade describes how the transmission is suppressed below these intrinsic values, and our protocol to extract values of intrinsic $\tau_{1,2}$ from conductance measurements.}

\subsection*{Dynamical Coulomb Blockade Framework}\label{sec:DCB}

\textcolor{black}{For arbitrary mesoscopic conductors embedded in a purely dissipative environment, the suppression in the zero-bias transmssion is given by,}

\begin{equation}
    \frac{\tilde{\tau}(V_{\text{dc}})}{1 - \tilde{\tau}(V_{\text{dc}})} = \frac{\tilde{\tau}(V_{\text{dc}}=0)}{1-\tilde{\tau}(V_{\text{dc}}=0)}\frac{E_B(R_{\text{env}}, V_{\text{dc}}, T_e) + 1}{E_B(R_{\text{env}}, V_{\text{dc}}=0, T_e) + 1},
    \label{eq:DCBmain}
\end{equation}

\textcolor{black}{where $\tilde{\tau}(V_{\text{dc}}) = h/e^2(G^{-1} - R_{\text{env}})^{-1}$ is the transmission extracted from a series conductance measurement at a bias voltage $V_{\text{dc}}$, and $E_B(R_{\text{env}}, V_{\text{dc}}=0, T_e):=\lim_{\tau \to 0} \left(\tilde{\tau}(V_{\text{dc}}=0) - \tau\right)/\tau$ is the relative circuit backaction for a tunnel junction in the same circuit. The backaction $E_B(R, V_{\text{dc}}, T_e)$ for arbitrary bias voltages can be computed using the dynamical Coulomb blockade framework for tunnel junctions. }
We review the expressions for conductance suppression of a short conductor in the weak tunneling limit $(\tau \ll 1)$ embedded in a resistive environment, as calculated by Joyez and Esteve~\cite{Joyez1998}.

Considering the environment impedance $Z$, temperature $T$, and bias voltage $V$, the differential conductance $G = dI/dV$ through the short conductor is given by 
\begin{equation}
\begin{split}
G(V)
&= \frac{1}{R_T}\Biggl[
1 + 2 \int_0^\infty \! dt\,
\pi t \left(\frac{k_B T_e}{\hbar}\right)^2 \\[2pt]
&\quad\times
\Im\!\left[e^{J(t)}\right]
\cos\!\left(\frac{e V t}{\hbar}\right)
\sinh^{-2}\!\left(\frac{\pi t k_B T_e}{\hbar}\right)
\Biggr].
\end{split}
\end{equation}
        
% \end{align}
where $R_T = 1/G_{\infty}$ is the intrinsic tunnel resistance, taken to be much larger than the environment resistance in this framework $\Re[Z] = R_{\text{env}} \ll R_T$.

Modeling the environment as an $RC$ circuit with $R_{\text{env}} = \frac{h}{e^2\tau_2}$ set by the top QPC transmission and $C = e^2/2E_c$ determined by the hybrid dot charging energy, $J(t)$ is given by 

\begin{equation}
\begin{split}
J(t)
&= \frac{\pi R_{\text{env}}}{R_K}\Biggl[
\left(1 - e^{-|t|/R_{\text{env}}C}\right)
\left(\cot\!\frac{\hbar}{2R_{\text{env}}Ck_B T_e} - i\right) \\[2pt]
&\quad - \frac{2 k_B T_e |t|}{\hbar}
+ 2 \sum_{n=1}^{\infty}
\frac{1 - e^{-\omega_n |t|}}
{2\pi n\!\left[(R_{\text{env}}C\omega_n)^2 - 1\right]}
\Biggr].
\end{split}
\end{equation}

where $\omega_n = \frac{2\pi n k_BT}{\hbar}$ are the Matsubara frequencies for $n\in \mathbb{Z}^+$, and 

\begin{equation}
\begin{split}
2 \sum_{n=1}^{\infty}
\frac{1 - e^{-\omega_n |t|}}
{2\pi n\!\left[(R_{\text{env}}C\omega_n)^2 - 1\right]}
&= - \frac{1}{\pi}\Biggl[
2\gamma + \Psi(-x) + \Psi(x) \\[2pt]
&\quad + 2\ln(1-y)
+ \frac{y}{1+x} \\[2pt]
&\qquad {}\times
{}_2F_1(1,1+x,2+x,y) \\[4pt]
&\quad + \frac{y}{1-x} \\[2pt]
&\qquad {}\times
{}_2F_1(1,1-x,2-x,y)
\Biggr].
\end{split}
\end{equation}

where $\gamma$ is Euler's constant, $\Psi$ is the logarithmic derivative of the Gamma function, $_2F_1$ is the hypergeometric function, $y = \exp{(\frac{-2\pi t k_B T_e}{\hbar})}$, and $x = \frac{h}{e^2}\frac{E_c}{4 \pi^2 R_{\text{env}}k_BT}$.

\section*{Data Availability Statement}
All data used in this work are available in the Stanford Digital Repository at https://doi.org/10.25740/nh213xs3377.

\section*{Acknowledgments}

The authors thank B. Suo, W. Pouse, M. Pendharkar, C. Marcus, C. Mora, F. Pierre, A. Anthore, H. Sahasrabudhe, and R. Rahman for their scientific insights and suggestions. Measurement and analysis were primarily supported by the U.S. Department of Energy (DOE), Office of Science, Basic Energy Sciences (BES), under Contract No. DE-AC02-76SF00515. The Gordon \& Betty Moore Foundation supported infrastructure under Grant No.\ GBMF9460, as well as completion of the project. C.L.H. acknowledges fellowship support from the National Science Foundation (NSF) and Stanford Graduate Fellowship (SGF). K.A.M. acknowledges fellowship support from the National Science Foundation (NSF). Part of this work was performed at nano@stanford RRID:SCR\_026695. E.S. gratefully acknowledges support from the European Research Council (ERC) under the European Union Horizon 2020 research and innovation programme under grant agreement No. 951541.  A.K.M acknowledges funding from Research Ireland through grant
24/FFP-P/12816.

\section*{Author Contributions}
P.S., C.L.H., and K.A.M. performed the measurements. C.L.H. fabricated the device.  
P.S., C.L.H., K.A.M., M.A.K., and D.G.-G. analyzed the data.  
T.W., C.T., G.C.G., and M.J.M. grew the InAs heterostructure.
A.K.M. and E.S. developed the quantum simulation proposal.
P.S. wrote the manuscript with contributions from K.A.M., A.K.M., E.S., D.G.-G. and feedback from all coauthors. D.G.-G. supervised the project.

\section*{Competing Interests}
The authors declare no competing financial or non-financial interests.

\section*{Supplementary Information}
The supplemental material document contains details on the (1) Landauer-B\"{u}ttiker analysis for device conductance and edge mode transparency (2) magnetic field-dependent equilibration of quantum Hall edge modes, (3) charge noise spectroscopy measurement setup, (4) sensitivity of the series conductance to plunger and quantum point contact gate voltages, (5) analysis of the charge noise contribution to Coulomb peak width, (6) charge quantization measurements in the nearly-ballistic limit (7) Coulomb diamond and electron thermometry data for a different hybrid dot (Device C), (8) dynamical Coulomb blockade in the ballistic limit in Device C, (9) intrinsic transmission estimations for Device B, and (10) Hamiltonian of the hybrid dot array 
%%%%%%% MAIN TEXT BIB START %%%%%%%%%%%
%apsrev4-2.bst 2019-01-14 (MD) hand-edited version of apsrev4-1.bst
%Control: key (0)
%Control: author (8) initials jnrlst
%Control: editor formatted (1) identically to author
%Control: production of article title (0) allowed
%Control: page (0) single
%Control: year (1) truncated
%Control: production of eprint (0) enabled
\providecommand{\noopsort}[1]{}\providecommand{\singleletter}[1]{#1}%
\pagebreak
\widetext
\begin{center}
\textbf{\large Supplemental Material: Hybrid metal-semiconductor quantum dots in InAs as a platform for quantum simulation}
\end{center}
\date{\today}
%%%%%%%%%% Merge with supplemental materials %%%%%%%%%%
%%%%%%%%%% Prefix a "S" to all equations, figures, tables and reset the counter %%%%%%%%%%
\setcounter{equation}{0}
\setcounter{figure}{0}
\setcounter{table}{0}
\setcounter{page}{1}
\makeatletter
\renewcommand{\theequation}{S\arabic{equation}}
\renewcommand{\thefigure}{S\arabic{figure}}
\renewcommand{\thesection}{S\arabic{section}}
\renewcommand{\bibnumfmt}[1]{[S#1]}

\section{Landauer-B\"{u}ttiker Analysis for device conductance and edge mode transparency}

\subsection{Derivation of Device Conductance in Quantum Hall}

In the following section, we derive equations \ref{eq:QHbuttikerconductance} and \ref{eq:transparencyouterLB} in the main text using a Landauer Buttiker analysis. To simplify our analysis, we use a modified diagram shown below in Figure \ref{fig:LBdiagram}, whose layout of ohmics is identical to the device shown in Figure \ref{fig:transparency} of the main text, but which has the ohmics labelled for convenience. 

\FloatBarrier
\begin{figure}[!htbp] 
        %\begin{minipage}[b]{0.45\linewidth}
            \centering\includegraphics[width=0.8\textwidth,keepaspectratio]{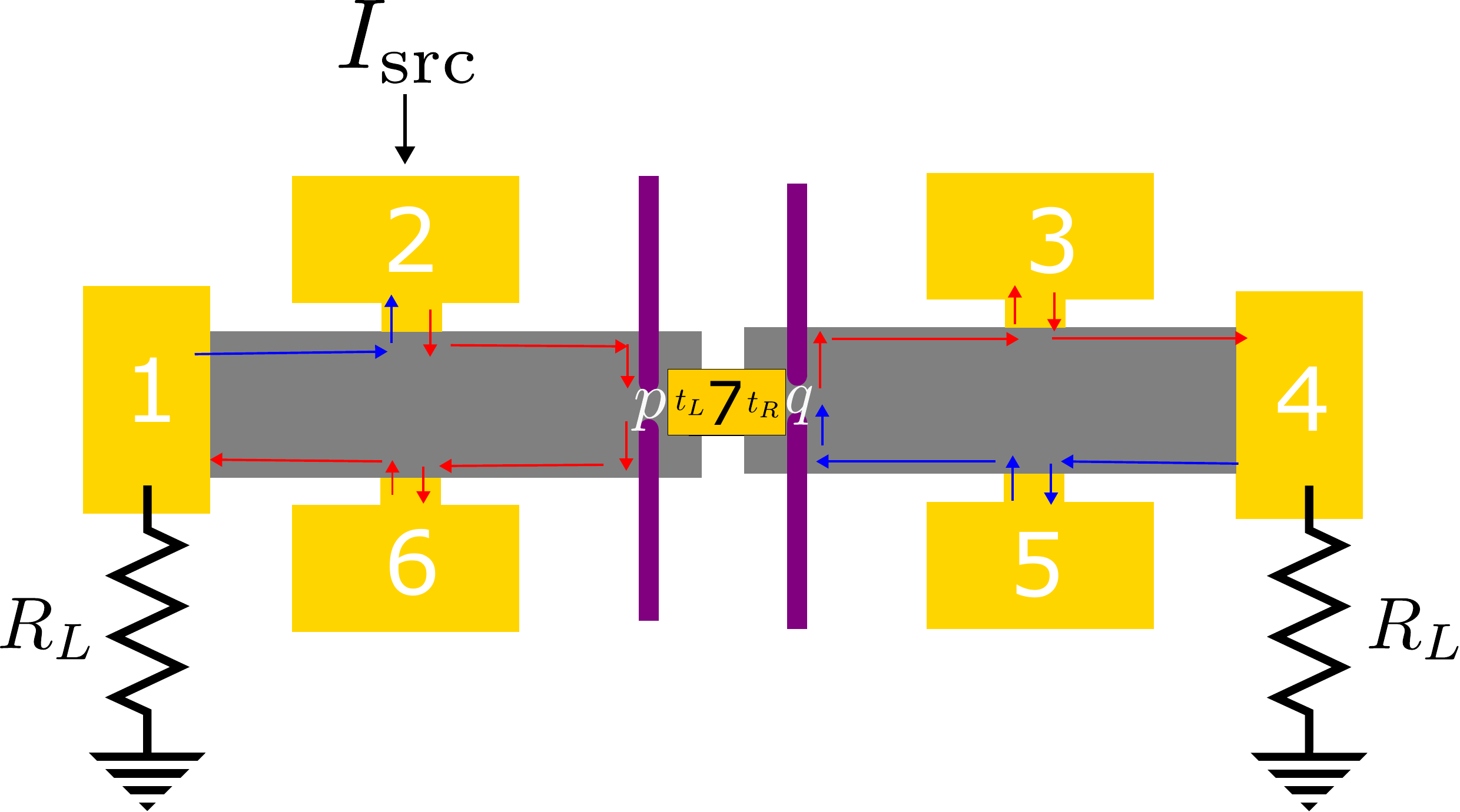}
\caption{\textbf{Edge mode transparency device schematic}. The schematic represents a Hall bar-like mesa (gray) with a trench in the middle. The metal island \{7\} makes ohmic contact to the 2DEG on either side of the trench, with average transparency $t_L$ $(t_R)$ to the edge modes on the left (right). Macroscopic ohmic leads \{1, $\cdots$, 6\} are used for transport measurements, and narrow QPC gate electrodes transmit a fraction $p$ ($q$) of the quantum Hall edge modes at the left (right) side of the island. Current $I_{src}$ is sourced at 2, and drained at 1 and 4 via a line resistance $R_L$. A clockwise chirality of the edge states is assumed.} 
 	 \label{fig:LBdiagram}
        %\end{minipage}
      % \hspace{0.5cm}
        %\pause
            \end{figure}

For a derivation of equation \ref{eq:QHbuttikerconductance}, we begin by neglecting the details of the edge state transparency through the island, and derive the conductance through the entire device as a function of the voltages measured on each terminal.

The following calculation is done based on the formalism described in Ref.~\cite{datta_mesoscopic}.

\begin{equation}
G_{x\leftarrow y} = G_0 \times
\begin{blockarray}{ccccccccc}
&&& 1 & 2 & 3 & 4 & 5 & 6  \\
\begin{block}{cc(ccccccc)}
  1 &&& 0 & 0 & 0 & 0 & 0 & 1  \\
  2 &&& 1 & 0 & 0 & 0 & 0 & 0 \\
  3 &&& 0 & G_{32} & 0 & 0 & G_{35} & 0  \\
  4 &&& 0 & 0 & 1 & 0 & 0 & 0 \\
  5 &&& 0 & 0 & 0 & 1 & 0 & 0 \\
  6 &&& 0 & G_{62} & 0 & 0 & G_{65}& 0 \\
\end{block}
\end{blockarray}\hspace{0.2cm},
\end{equation}
Here, we note that $G_0 = \nu\frac{e^2}{h}$ is the longitudinal quantum hall conductance, and
\begin{align*}
    G_{32} &= \frac{G_{\rm{device}}}{G_0} = \alpha\\
    G_{35} &= \frac{G_0-G_{\rm{device}}}{G_0}  = 1- \alpha \\
    G_{62} &=  \frac{G_0-G_{\rm{device}}}{G_0}  = 1- \alpha \\
    G_{65} &= \frac{G_{\rm{device}}}{G_0} = \alpha\\
\end{align*}
\noindent
These coefficients $G_{32}$ and $G_{65}$ represent the device conductance, whereas the other formulas can be derived from properties of the conductance matrix $\sum_{y} G_{x\leftarrow y} = G_0$. We now can invoke the Buttiker formula
\begin{equation}
\label{eq:Buttiker}
    I_x = \sum_y\left[G_{yx}V_x - G_{xy}V_y\right]
\end{equation}

In addition to $\sum_{y} G_{x\leftarrow y} = G_0$, we also know $\sum_{x} G_{x\leftarrow y} = G_0$, so we can write
\begin{align*}
 I_x & = \sum_y\left[G_{yx}V_x - G_{xy}V_y\right]\\ 
    I_x &= G_0 V_x -\sum_y\left[ G_{xy}V_y\right]\\
     &= \sum_y\left[ (G_0 \delta_{xy} - G_{xy})V_y\right]
\end{align*}

\begin{equation}
\begin{pmatrix}
I_1 \\ I_2 \\ I_3 \\ I_4 \\ I_5 \\ I_6 
\end{pmatrix} = G_0 \begin{pmatrix}
 1 & 0 & 0 & 0 & 0 & -1  \\
-1 & 1 & 0 & 0 & 0 & 0 \\
 0 & -\alpha & 1 & 0 & \alpha-1 & 0  \\
 0 & 0 & -1 & 1 & 0 & 0 \\
 0 & 0 & 0 & -1 & 1& 0 \\
0 & \alpha-1 & 0 & 0 & -\alpha& 1 \\
\end{pmatrix}
\begin{pmatrix}
V_1 \\ V_2 \\ V_3 \\ V_4 \\ V_5 \\ V_6 
\end{pmatrix}
\end{equation}

We further note that 

\begin{equation*}
\begin{pmatrix}
I_1 \\ I_2 \\ I_3 \\ I_4 \\ I_5 \\ I_6 
\end{pmatrix} = 
\begin{pmatrix}
-V_1/R_L\\ I_{\rm{src}} \\ 0 \\ -V_4/R_L\\ 0 \\ 0
\end{pmatrix}  =  \begin{pmatrix}
-G_0 \frac{V_1}{\delta} \\ I_{\rm{src}} \\ 0 \\ -G_0 \frac{V_4}{\delta}\\ 0 \\ 0
\end{pmatrix},
\end{equation*}
since contacts 1 and 4 are grounded with line resistance $R_L$, contact 2 has a source current $I_{\rm{src}}$, and all other contacts are floating. Here, $\delta = R_L G_0$ is a dimensionless line resistance. We have

\begin{align*}
-G_0 \begin{pmatrix}
 \frac{1}{\delta} & 0 & 0 & 0 & 0 & 0  \\
0 & 0 & 0 & 0 & 0 & 0 \\
 0 & 0 & 0 & 0 & 0 & 0  \\
 0 & 0 & 0 & \frac{1}{\delta} & 0 & 0 \\
 0 & 0 & 0 & 0 & 0& 0 \\
0 & 0 & 0 & 0 & 0 & 0 \\
\end{pmatrix}\begin{pmatrix}
V_1\\ V_2 \\ V_3 \\ V_4 \\ V_5 \\ V_6 
\end{pmatrix} + \begin{pmatrix}
0\\ I_{\rm{src}} \\ 0 \\ 0 \\ 0\\ 0 
\end{pmatrix}  &= G_0 \begin{pmatrix}
 1 & 0 & 0 & 0 & 0 & -1  \\
-1 & 1 & 0 & 0 & 0 & 0 \\
 0 & -\alpha & 1 & 0 & \alpha -1& 0  \\
 0 & 0 & -1 & 1 & 0 & 0 \\
 0 & 0 & 0 & -1 & 1& 0 \\
0 & \alpha-1 & 0 & 0 & -\alpha& 1 \\
\end{pmatrix}
\begin{pmatrix}
V_1 \\ V_2 \\ V_3 \\ V_4 \\ V_5 \\ V_6 
\end{pmatrix}\\
\begin{pmatrix}
0\\ I_{\rm{src}} \\ 0 \\ 0 \\ 0\\ 0 
\end{pmatrix}  &= G_0 \begin{pmatrix}
 1 + \frac{1}{\delta} & 0 & 0 & 0 & 0 & -1  \\
-1 & 1 & 0 & 0 & 0 & 0 \\
 0 & -\alpha & 1 & 0 & \alpha-1 & 0  \\
 0 & 0 & -1 & 1 + \frac{1}{\delta} & 0 & 0 \\
 0 & 0 & 0 & -1 & 1& 0 \\
0 & \alpha-1 & 0 & 0 & -\alpha& 1 \\
\end{pmatrix}
\begin{pmatrix}
V_1 \\ V_2 \\ V_3 \\ V_4 \\ V_5 \\ V_6 
\end{pmatrix} \\
G_0^{-1} \begin{pmatrix}
 1 + \frac{1}{\delta} & 0 & 0 & 0 & 0 & -1  \\
-1 & 1 & 0 & 0 & 0 & 0 \\
 0 & -\alpha & 1 & 0 & \alpha-1 & 0  \\
 0 & 0 & -1 & 1 + \frac{1}{\delta} & 0 & 0 \\
 0 & 0 & 0 & -1 & 1& 0 \\
0 & \alpha-1 & 0 & 0 & -\alpha& 1 \\
\end{pmatrix}^{-1}
\begin{pmatrix}
0 \\ I_{\rm{src}}  \\ 0 \\ 0 \\ 0 \\ 0
\end{pmatrix} & = \begin{pmatrix}
V_1 \\ V_2 \\ V_3 \\ V_4 \\ V_5 \\ V_6 
\end{pmatrix} 
\end{align*}
The solutions are, for the case of $\delta = 0$, are given by
\begin{alignat*}{4}
    V_1 &= \frac{\delta + \alpha(\delta -1) \delta}{G_0 (1+ 2\alpha \delta )} I_{\rm{src}} &&= 0 \\
    V_2 &= \frac{(1+\delta)(1+\alpha \delta)}{G_0 (1+ 2\alpha \delta )} I_{\rm{src}} &&= \frac{I_{\rm{src}}}{G_0}\\ 
    V_3 &= \frac{\alpha(1+\delta)^2}{G_0 (1+ 2\alpha \delta )} I_{\rm{src}} &&= \frac{\alpha I_{\rm{src}}}{G_0}\\
    V_4 &= \frac{\alpha \delta(1+\delta)}{G_0 (1+ 2\alpha \delta )}I_{\rm{src}} && = 0\\
    V_5 &= V_4  &&= 0\\
    V_6 &= \frac{1-\alpha + \delta +  \alpha \delta^2}{G_0 (1+ 2\alpha \delta )} I_{\rm{src}} &&= 
 \frac{(1-\alpha)I_{\rm{src}}}{G_0} \\
\end{alignat*}
Thus, 
\begin{equation*}
    \alpha = \frac{V_3}{V_3 + V_6}
\end{equation*}
remembering that $\alpha =  \frac{G_{\rm{device}}}{G_0}$, with $G_0  = \nu \frac{e^2}{h}$, identifying the transmitted voltage $V_T = V_3$, and reflected voltage $V_R = V_6$, we arrive at equation \ref{eq:QHbuttikerconductance} from the main text
\begin{equation*}
    G_{\rm{device}} = \nu \frac{e^2}{h} \frac{V_T}{V_T + V_R}.
\end{equation*}
\subsection{Landauer Buttiker analysis of island transparency}
We now perform an Landauer Buttiker analysis to determine the transparencies of the small metal island in ohmic contact with the semiconductor. Because the island acts as a small ohmic contact, the full treatment that allows us to separate independently the right and left interfaces necessitates a 7x7 matrix, instead of the 6x6 matrix above.
\begin{equation}
G_{x\leftarrow y} = G_0 \times
\begin{blockarray}{cccccccccc}
&&& 1 & 2 & 3 & 4 & 5 & 6 & 7 \\
\begin{block}{cc(cccccccc)}
  1 &&& 0 & 0 & 0 & 0 & 0 & 1 & 0 \\
  2 &&& 1 & 0 & 0 & 0 & 0 & 0 & 0\\
  3 &&& 0 & 0 & 0 & 0 & G_{35} & 0 & G_{37} \\
  4 &&& 0 & 0 & 1 & 0 & 0 & 0 & 0\\
  5 &&& 0 & 0 & 0 & 1 & 0 & 0 & 0\\
  6 &&& 0 & G_{62} & 0 & 0 & 0 & 0 & G_{67}\\
  7 &&& 0 & G_{72} & 0 & 0 & G_{75} & 0 & G_{77}\\
\end{block}
\end{blockarray}\hspace{0.2cm},
\end{equation}
 
where $G_0 = \nu_0 e^2/h$ for bulk filling factor $\nu_0$, and 
\begin{align}
    G_{35} &= (1-q) + q(1-t_R) = 1-qt_R\\
    G_{37} &= qt_R\\
    G_{62} &= (1-p) + p(1-t_L) = 1-pt_L\\
    G_{67} &= pt_L\\
    G_{72} &= pt_L\\
    G_{75} &= qt_R
\end{align}

 Here $p,q = \nu_{1,2}/\nu_0$ is the ratio of local filling factor to bulk filling factor in QPCs 1 and 2 respectively, and $t_{L,R}$ is the mean transparency of the island to edge states on the left and right side respectively.
%  Note that I don't care about $G_{77}$ since it won't play a role in the Buttiker formula (see Eq.~\ref{eq:Buttiker}).
We can invoke equation \ref{eq:Buttiker} and a similar analysis to above, we have
\begin{equation*}
\label{eq:Buttiker}
    I_x = \sum_y\left[G_{yx}V_x - G_{xy}V_y\right]
\end{equation*}
\begin{align*}
\label{eq:IV}
\begin{pmatrix}
I_1 \\ I_2 \\ 0 \\ I_4 \\ 0 \\ 0 \\ 0
\end{pmatrix} &= G_0 \times
\begin{pmatrix}
   1 & 0 & 0 & 0 & 0 & -1 & 0 \\
  -1 & 1 & 0 & 0 & 0 & 0 & 0\\
  0 & 0 & 1 & 0 & -G_{35} & 0 & -G_{37} \\
   0 & 0 & -1 & 1 & 0 & 0 & 0\\
  0 & 0 & 0 & -1 & 1 & 0 & 0\\
  0 & -G_{62} & 0 & 0 & 0 & 1 & -G_{67}\\
  0 & -G_{72} & 0 & 0 & -G_{75} & 0 & G_{72} + G_{75}\\
\end{pmatrix}
\begin{pmatrix}
V_1 \\ V_2 \\ V_3 \\ V_4 \\ V_5 \\ V_6 \\ V_7
\end{pmatrix}\\
 &= G_0 \times
\begin{pmatrix}
   1 & 0 & 0 & 0 & 0 & -1 & 0 \\
  -1 & 1 & 0 & 0 & 0 & 0 & 0\\
  0 & 0 & 1 & 0 & q t_R - 1 & 0 & -qt_R \\
   0 & 0 & -1 & 1 & 0 & 0 & 0\\
  0 & 0 & 0 & -1 & 1 & 0 & 0\\
  0 & pt_L-1 & 0 & 0 & 0 & 1 & -pt_L\\
  0 & - pt_L & 0 & 0 & -qt_R & 0 & pt_L + qt_R\\
\end{pmatrix}
\begin{pmatrix}
V_1 \\ V_2 \\ V_3 \\ V_4 \\ V_5 \\ V_6 \\ V_7
\end{pmatrix},
\end{align*}
 where again we have used $I_{3, 5, 6, 7} = 0$ for floating contacts. Similarly to above, we can  relate the current at the drains to the dimensionless line resistance $\delta = R_L G_0$
 \begin{equation}
 \label{eq:Id}
     I_{1,4} = -V_{1,4}/R_L = -G_0 \frac{V_{1,4}}{\delta},
 \end{equation}
 where the current is positive for current flowing into the device. Just as above, we have a current $I_{\rm{src}}$ sourced at contact 2. 

 \begin{align*}
-G_0 \begin{pmatrix}
 \frac{1}{\delta} & 0 & 0 & 0 & 0 & 0  \\
0 & 0 & 0 & 0 & 0 & 0 \\
 0 & 0 & 0 & 0 & 0 & 0  \\
 0 & 0 & 0 & \frac{1}{\delta} & 0 & 0 \\
 0 & 0 & 0 & 0 & 0& 0 \\
0 & 0 & 0 & 0 & 0 & 0 \\
\end{pmatrix}\begin{pmatrix}
V_1\\ V_2 \\ V_3 \\ V_4 \\ V_5 \\ V_6 
\end{pmatrix} + \begin{pmatrix}
0\\ I_{\rm{src}} \\ 0 \\ 0 \\ 0\\ 0 
\end{pmatrix}  &= G_0
\begin{pmatrix}
   1 & 0 & 0 & 0 & 0 & -1 & 0 \\
  -1 & 1 & 0 & 0 & 0 & 0 & 0\\
  0 & 0 & 1 & 0 & q t_R - 1 & 0 & -qt_R \\
   0 & 0 & -1 & 1 & 0 & 0 & 0\\
  0 & 0 & 0 & -1 & 1 & 0 & 0\\
  0 & pt_L-1 & 0 & 0 & 0 & 1 & -pt_L\\
  0 & - pt_L & 0 & 0 & -qt_R & 0 & pt_L + qt_R\\
\end{pmatrix}
\begin{pmatrix}
V_1 \\ V_2 \\ V_3 \\ V_4 \\ V_5 \\ V_6 
\end{pmatrix}\\
\begin{pmatrix}
0\\ I_{\rm{src}} \\ 0 \\ 0 \\ 0\\ 0 
\end{pmatrix}  &= G_0 
\begin{pmatrix}
   1 + \frac{1}{\delta} & 0 & 0 & 0 & 0 & -1 & 0 \\
  -1 & 1 & 0 & 0 & 0 & 0 & 0\\
  0 & 0 & 1 & 0 & q t_R - 1 & 0 & -qt_R \\
   0 & 0 & -1 & 1+\frac{1}{\delta} & 0 & 0 & 0\\
  0 & 0 & 0 & -1 & 1 & 0 & 0\\
  0 & pt_L-1 & 0 & 0 & 0 & 1 & -pt_L\\
  0 & - pt_L & 0 & 0 & -qt_R & 0 & pt_L + qt_R\\
\end{pmatrix}
\begin{pmatrix}
V_1 \\ V_2 \\ V_3 \\ V_4 \\ V_5 \\ V_6 
\end{pmatrix} \\
\begin{pmatrix}
V_1 \\ V_2 \\ V_3 \\ V_4 \\ V_5 \\ V_6 
\end{pmatrix} &= 
G_0^{-1} \begin{pmatrix}
   1 + \frac{1}{\delta} & 0 & 0 & 0 & 0 & -1 & 0 \\
  -1 & 1 & 0 & 0 & 0 & 0 & 0\\
  0 & 0 & 1 & 0 & q t_R - 1 & 0 & -qt_R \\
   0 & 0 & -1 & 1+\frac{1}{\delta} & 0 & 0 & 0\\
  0 & 0 & 0 & -1 & 1 & 0 & 0\\
  0 & pt_L-1 & 0 & 0 & 0 & 1 & -pt_L\\
  0 & - pt_L & 0 & 0 & -qt_R & 0 & pt_L + qt_R\\
\end{pmatrix}^{-1} \begin{pmatrix}
0\\ I_{\rm{src}} \\ 0 \\ 0 \\ 0\\ 0 
\end{pmatrix} 
\end{align*}

\begin{align*}
     V_1 &= \frac{\delta(qt_R + p(t_L + (\delta - 1)qt_Lt_R))}{G_0(qt_R + p(t_L + 2\delta qt_Lt_R))}I_{\rm{src}}\\
    V_2 &= \frac{(\delta + 1)(qt_R + p(t_L + \delta qt_Lt_R))}{G_0(qt_R + p(t_L + 2\delta qt_Lt_R))}I_{\rm{src}}\\
    V_3 &= \frac{(1+\delta)^2pqt_Lt_R}{G_0(qt_R + p(t_L + 2\delta qt_Lt_R))}I_{\rm{src}}\\
    V_4 &= \frac{\delta(1+\delta)pqt_Lt_R}{G_0(qt_R + p(t_L + 2\delta qt_Lt_R))}I_{\rm{src}}\\
    V_5 &= V_4\\
    V_6 &= \frac{(1+\delta)(qt_R + p(t_L + (\delta - 1)qt_Lt_R))}{G_0(qt_R + p(t_L + 2\delta qt_Lt_R))}I_{\rm{src}}\\
    V_7 &= \frac{(1+\delta)pt_L(1+\delta qt_R)}{G_0(qt_R + p(t_L + 2\delta qt_Rt_R))}I_{\rm{src}}
\end{align*}
If we consider the case of zero line resistance $\delta=0$, we have
\begin{align*}
    V_2 &= \frac{I_{\rm{src}}}{G_0}\\
    V_1 &= 0\\
    V_3 &= \frac{pqt_Lt_R}{G_0(qt_R + pt_L) }I_{\rm{src}}\\
    V_4 &= 0\\
    V_5 &= V_4\\
    V_6 &= \frac{(qt_R + pt_L   - pqt_Lt_R)}{G_0(qt_R + pt_L)}I_{\rm{src}}\\
    V_7 &= \frac{pt_L}{G_0(qt_R + pt_L)}I_{\rm{src}}
\end{align*}
Below, we show a table showing the transmitted voltage ratio $V_3/V_2$ as a function of the filling factor ratio in each QPC $q, p$ as a function of the transparency (assuming symmetric interfaces such that $t_L = t_R = t$)
\begin{equation}
    V_3/V_2 = 
\begin{array}{|c||c|c|c|c|c|c|c|}
    \hline
    q \downarrow, p \rightarrow & 0 & 1/4 & 1/3 & 1/2 & 2/3 & 3/4 & 1 \\
    \hline
    \hline
    0 & 0 & 0 & 0 & 0 & 0 & 0 & 0 \\
    \hline
    1/4 & 0 & t/8 & - & t/6 & - & 3t/16 & t/5 \\
    \hline
    1/3 & 0 & - & t/6 & - & 2t/9 & - & t/4 \\
    \hline
    1/2 & 0 & t/6 & - & t/4 & - & 3t/10 & t/3 \\
    \hline
    2/3 & 0 & - & 2t/9 & - & t/3 & - & 2t/5 \\
    \hline
    3/4 & 0 & 3t/16 & - & 3t/10 & - & 3t/8 & 3t/7 \\
    \hline
    1 & 0 & t/5 & t/4 & t/3 & 2t/5 & 3t/7 & t/2 \\
    \hline
    
    \hline
    \hline

\end{array}
\end{equation}

\begin{equation}
V_6/V_2 = 1 - V_3/V_2
\end{equation}

\begin{equation}
V_7/V_2 = 
\begin{array}{|c||c|c|c|c|c|c|c|}
    \hline
    q \downarrow, p \rightarrow & 0 & 1/4 & 1/3 & 1/2 & 2/3 & 3/4 & 1 \\
    \hline
    \hline
    0 & - & 1 & 1 & 1 & 1 & 1 & 1 \\
    \hline
    1/4 & 0 & 1/2 & - & 2/3 & - & 3/4 & 4/5 \\
    \hline
    1/3 & 0 & - & 1/2 & - & 2/3 & - & 3/4 \\
    \hline
    1/2 & 0 & 1/3 & - & 1/2 & - & 3/5 & 2/3 \\
    \hline
    2/3 & 0 & - & 1/3 & - & 1/2 & - & 3/5 \\
    \hline
    3/4 & 0 & 1/4 & - & 2/5 & - & 1/2 & 4/7 \\
    \hline
    1 & 0 & 1/5 & 1/4 & 1/3 & 2/5 & 3/7 & 1/2 \\
    \hline
    
    \hline
    \hline
\end{array}
\end{equation}

\subsection{Edge-state dependent transparency solutions}
Especially at low quantum hall filling factors, when there can be a large spatial distributions of edge modes, the transparencies of the innermost quantum hall edge mode can often be significantly different from the outermost edge modes. 

In order to calculate the individual transparencies of the edge modes, we can use the following prescription. When a QPC has filling factor ratio $q,p = \frac{k}{\nu}$, where $k$ is the number of edge modes let through and $\nu$ is the bulk filling factor, we can replace
\begin{equation*}
    p t_L \to  p\overline{t_{\nu-k+1, L}}, q t_R \to  q\overline{t_{\nu-k+1, R}}
\end{equation*}
with
\begin{align*}
    \overline{t_{\nu-k+1, L}} &= \frac{1}{k}\sum_{n=1}^{k} t_{ n,L}\\
    \overline{t_{\nu-k+1, R}} &= \frac{1}{k}\sum_{n=1}^{k} t_{n, R}
\end{align*}
i.e. we let $k$ edge modes through the QPC, the transparency used in the Landauer-Buttiker formulas is the average of the $k$ outer-most edge state transparencies. 

If we set each QPC to transmit a single quantum hall edge mode (assuming symmetric interfaces such that $t_{\nu, L} = t_{\nu, R} = t_\nu$), we derive equation \ref{eq:transparencyouterLB} in the main text
\begin{align*}
    \frac{V_6}{V_2} &= \frac{V_R^{\tau_{1,2}=1}}{V_R^{\tau_{1,2}=0}} \\
    & =  \frac{qt_{\nu, R} + pt_{\nu,L}   - pqt_{\nu,L}t_{\nu,R}}{qt_{\nu,R} + pt_{\nu, L}} \\
    & =  \frac{t_\nu/\nu + t_\nu/\nu   - t_\nu^2/\nu^2}{t_\nu/\nu + t_\nu/\nu} \\
     & =  1 - \frac{t_\nu}{2\nu}.
\end{align*}

A more complicated case exists if we consider more than a single edge mode being allowed in one or both of the QPC. For example, if $p = \frac{1}{3}$ and $q = \frac{2}{3}$, with $\nu =3$, we have

\begin{align*}
    \frac{V_3}{V_2} &= \left(\frac{1}{3}\right)\left(\frac{2}{3}\right)\frac{t_{3, L}\overline{t_{2, R}}}{(\frac{2}{3}\overline{t_{2,R}} + \frac{1}{3}t_{3, L}) }\\
    &= \frac{2}{3}\frac{t_{3, L}\overline{t_{2, R}}}{(2\overline{t_{2, R}} + t_{3,L}) }, \hspace{0.5 cm} \overline{t_{2, R}} = \frac{1}{2}(t_{3, R} + t_{2, R})
\end{align*}

Following this prescription, we display the results of similar calculations that yield edge-state dependent transparency solutions, for filling factors $\nu = 2$ and $\nu = 3$ as a function of the filling factor ratios in both QPCs $p$ and $q$.

\subsubsection{$\nu = 2$}

\begin{equation}
V_t/V_r^{\tau=0} = 
\begin{array}{|c||c|c|c|c|c|c|c|}
    \hline
    q \downarrow, p \rightarrow & 0 & 1/2  & 1 \\
    \hline
    \hline
    0 & 0 & 0 & 0 \\
    \hline
    1/2 & 0 & {t_{2,L}t_{2,R}}/{2\left(t_{2,L}+t_{2,R}\right)} & {\overline{t_{L}}{t_{2,R}}}/{\left(2\overline{t_{L}}+{t_{2,R}}\right)} \\
    \hline
    1 & 0 & {{t_{2,L}}\overline{t_{R}}}/{\left({t_{2,L}}+2\overline{t_{R}}\right)} & {\overline{t_{L}} \overline{t_{R}}}/{\left({\overline{t_{L}}}+\overline{t_{R}}\right)} \\
    \hline
    
    \hline
    \hline

\end{array},
\end{equation}

where
\begin{align*}
    \overline{t_L} &= (t_{1,L} + t_{2,L})/2\\
    \overline{t_R} &= (t_{1,R} + t_{2,R})/2\\
\end{align*}

Assuming {$t_L = t_R$}

\begin{equation}
V_t/V_s = 
\begin{array}{|c||c|c|c|c|c|c|c|}
    \hline
    q \downarrow, p \rightarrow & 0 & 1/2  & 1 \\
    \hline
    \hline
    0 & 0 & 0 & 0 \\
    \hline
    1/2 & 0 & {t_{2}}/4 & {\overline{t}{t_{2}}}/{\left(2\overline{t}+{t_{2}}\right)} \\
    \hline
    1 & 0 & {{t_{2}}\overline{t}}/{\left({t_{2}}+2\overline{t}\right)} & {\overline{t}/ 2} \\
    \hline
    
    \hline
    \hline

\end{array},
\end{equation}

where
\begin{align*}
    \overline{t} &= (t_{2} + t_{1})/2
\end{align*}

\subsubsection{$\nu = 3$}

\begin{equation}
V_3/V_2 = 
\begin{array}{|c||c|c|c|c|c|c|c|}
    \hline
    q \downarrow, p \rightarrow & 0 & 1/3 & 2/3  & 1 \\
    \hline
    \hline
    0 & 0 & 0 & 0 & 0\\
    \hline
    1/3 & 0 & {t_{3,L}t_{3,R}}/{3\left(t_{3,L}+t_{3,R}\right)} & {2\overline{t_{2,L}}{t_{3,R}}}/{3\left(2\overline{t_{2,L}}+{t_{3,R}}\right)} & {\overline{t_{L}}{t_{3,R}}}/{\left({3\overline{t_{L}}}+{t_{3,R}}\right)} \\
    \hline
    2/3 & 0 & {2{t_{3,L}}\overline{t_{2,R}}}/{3\left({t_{3,L}}+2\overline{t_{2,R}}\right)} & {2\overline{t_{2,L}} \overline{t_{2,R}}}/{3\left({\overline{t_{2,L}}}+\overline{t_{2,R}}\right)} & {2\overline{t_{L}} \overline{t_{2,R}}}/{\left({3\overline{t_{L}}}+2\overline{t_{2,R}}\right)} \\
    \hline
     1 & 0 & {{t_{3,L}}\overline{t_{R}}}/{\left({t_{3,L}}+3\overline{t_{R}}\right)} & {2\overline{t_{2,L}} \overline{t_{R}}}/{\left({2\overline{t_{2,L}}}+3\overline{t_{R}}\right)} & {\overline{t_{L}} \overline{t_{R}}}/{\left({\overline{t_{L}}}+\overline{t_{R}}\right)}\\
    \hline
    \hline

\end{array},
\end{equation}

where
\begin{align*}
    \overline{t_L} &= (t_{1,L} + t_{2,L} + t_{3,L})/3\\
    \overline{t_R} &= (t_{1,R} + t_{2,R} + t_{3,R})/3\\
    \overline{t_{2,L}} &= (t_{3,L} + t_{2,L})/2\\
    \overline{t_{2,R}} &= (t_{3,R} + t_{2,R})/2\\
\end{align*}

In order to read out the transparencies of each individual edge state, we perform a QPC-QPC megasweep, where we measure the transmitted and reflected voltages as a function of the two QPC voltages. When each QPC is transmitting 1 edge mode, we can read out the transparency of the outermost edge mode $t_\nu$. We can then open each QPC to two edge modes, and use our measured outermost edge transparency and the measure of the outer two most edge state average transparency $\overline{t_{\nu-1}}$ to infer the transparency $t_{\nu-1}$. Each data point in Fig.~\ref{fig:transparency}(d) in the main text represents a 2-dimensional dataset similar to what is shown in Figure \ref{fig:transparency}b, where the diagonal elements $p=q$ can be used to extract individual edge state transparencies.

\section{Field-Dependent Equilibration of Quantum Hall Edge Modes}
\begin{figure}[!htbp] 
        %\begin{minipage}[b]{0.45\linewidth}
            \centering\includegraphics[width=
            \textwidth,keepaspectratio]{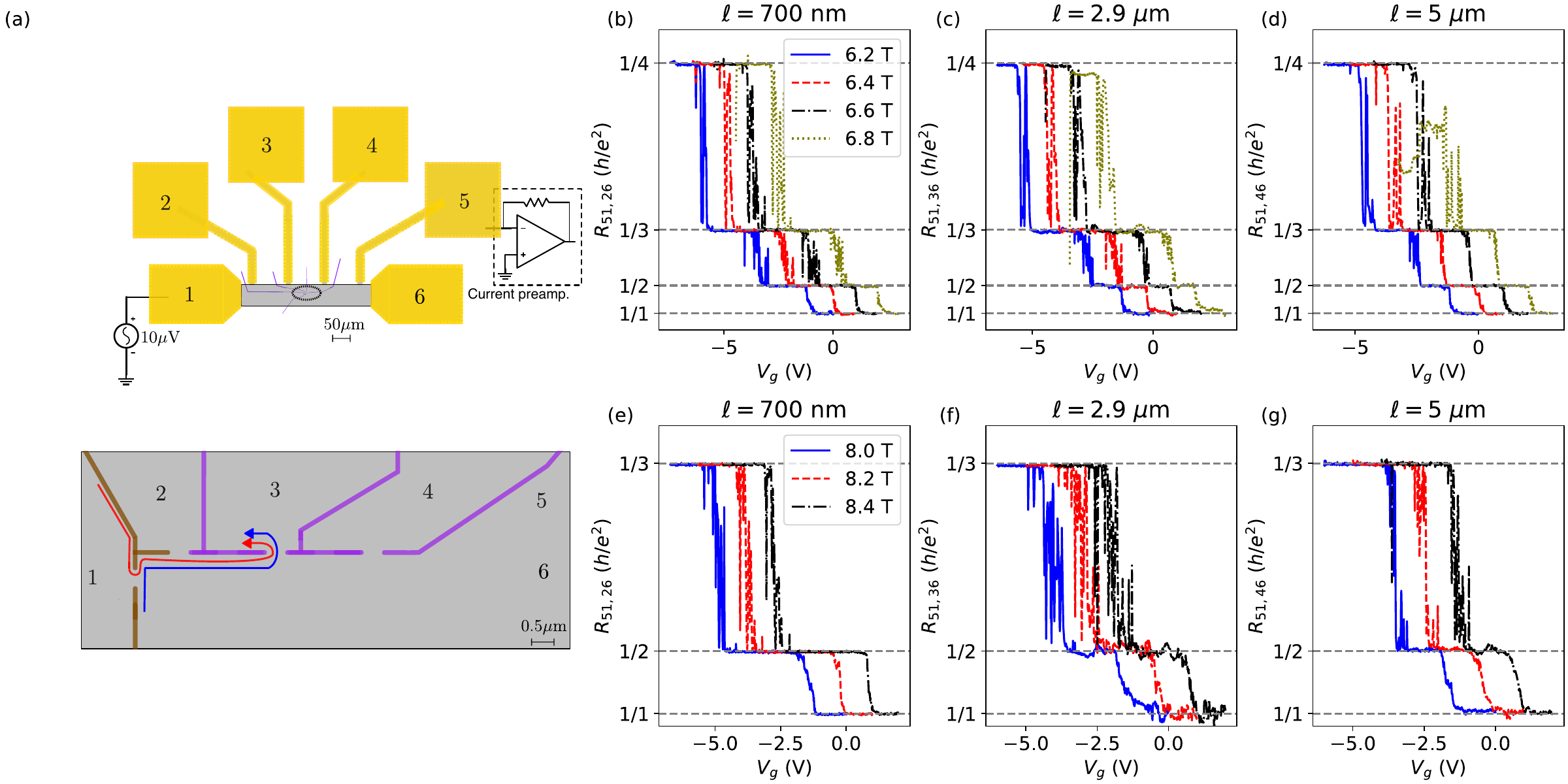}
\caption{\textbf{Quantum Hall equilibration length} (a) Device schematic for measuring quantum Hall equilibration and the measurement setup. The device features a 60 $\upmu$m-wide mesa (gray) with Ti/Au ohmic contacts (yellow), with quantum point contacts (dashed outline, expanded in the bottom panel) used for injecting and detecting a nonequilibrium distribution of chemical potentials. The injector (brown) is biased to the first conductance plateau to fully transmit the outermost edge mode, while the detectors (purple) are swept to measure the average chemical potential over an integer number of edge modes. The device is biased with a 10 $\upmu$V low-frequency ($<20$ Hz) ac excitation. The voltage drop between contacts $i$ and $6$ for $i\in \{2,3,4\}$ across the corresponding detector QPC is measured after differential amplification by a lock-in amplifier. The detector QPCs are placed $\ell = \{0.7, 2.9, 5\}~\upmu$m from the injector. A current preamplifier provides a virtual ground at the drain contact $5$, and a lock-in amplifier is used to measure the drain current. (b)-(d) Four-terminal resistance $R_{51,i6}$ for current flowing between $5\leftarrow 1$ and voltage drop across $i-6$ for $i \in \{2,3,4\}$ corresponding to $\ell = \{0.7, 2.9, 5\}~\upmu$m between the three panels for a set of four magnetic field settings on the $\nu = 4$ plateau. (e)-(g) Similar to (b)-(d), but for three magnetic field settings on the $\nu = 3$ plateau. } 
 	 \label{fig:QHeq}
 \end{figure}

The ability to inject a nonequlibrium distribution of edge mode chemical potentials through QPCs hinges on the assumption that the edge modes do not equilibrate over the $60~\upmu$m distance between the QPC and the island in Device A. We can test this assumption by studying the extent of equilibration as a function of distance and magnetic field with the device depicted in Fig.~\ref{fig:QHeq}(a). These measurements were done based on earlier work described in \cite{McEuen1991} and \cite{Alphenaar1991}. An \textit{injector} QPC, nominally identical to the QPC in Device A, is tuned to the first conductance plateau, corresponding to transmission of only the outermost edge mode. A set of three \textit{detector} QPCs are located adjacent to voltage probes at distances $\ell \in \{0.7, 2.9, 5\} \upmu $m from the injector, and are tuned to transmit  $n\in \mathbb{Z} \leq \nu$ edge modes.

Assuming no equilibration among the edge modes over the distance between the injector and the first detector with $n > 0$, a Landauer-B\"utticker treatment for chemical potential at the probe ``$i$'' behind the detector QPC reads

\begin{equation}
    \mu_{d} = \frac{1}{n} eV_S,  
    \label{eq:munoeq}
\end{equation}
where $eV_S$ is the chemical potential of the source contact ``1'' (see Fig.~\ref{fig:QHeq}), upstream the injector QPC. Similarly, in the limit where $m \leq \nu$ outer edge modes fully equilibrate before the detector QPC, $\mu_d$ is given by

\begin{equation}
    \mu_d = \begin{cases}
   (1/m) eV_S , & \text{if } n \leq m\\
   (1/n) eV_S, & \text{if } n > m.
\end{cases}
\label{eq:mueq}
\end{equation}

Comparing the voltage drop between the detector voltage probe ``$i$'' and the current drain ``6'', $eV_d = \mu_d$, with predictions from Eqs.~\ref{eq:munoeq},~\ref{eq:mueq} as a function of distance and magnetic field enables a quantitative study of edge mode equilibration. The measured four-terminal resistance $R_{51,i6}$ for current flowing between contacts $5\leftarrow1$ and voltage measured across $i-6$ where $i\in \{2,3,4\}$ is shown in Fig.~\ref{fig:QHeq}(b-g) together with predictions from Eqs.~\ref{eq:munoeq},~\ref{eq:mueq} for three magnetic fields each on the $\nu = 3,4$ quantum Hall plateaus. The measured $R_{51,i6}$ at $\nu=3$ is well explained by Eq.~\ref{eq:munoeq}, indicating no equilibration between the edge modes over the longest distance $\ell = 5~\upmu$m. In contrast, at the high field end of the $\nu=4$ quantum Hall plateau, $R_{51,36}$ and $R_{51,46}$ deviate from the predictions of Eq.~\ref{eq:munoeq} when the detector QPC is tuned to transmit only the outermost edge mode $(n=1)$. This indicates partial equilibration between the outer two edge modes over distances greater than 700 nm. Although we cannot predict the extent of equilibration over the $60~\upmu$m propagation length relevant for Device A, the results in this section indicate an equilibration length $l_{\text{eq},\nu}$ that reduces with increasing field for a given $\nu$, and is shorter for $\nu = 4$ (i.e., $l_{\text{eq},4} < l_{\text{eq},3}$). Further investigation of the $l_{\text{eq}, \nu}$ over longer injector-detector distances and higher filling factors is required to ascertain the role played by edge mode equilibration in the magnetic field and filling factor dependence of transparencies to submicron ohmic islands discussed in the transparency section of the main text.

\section{Details of Charge Noise Spectroscopy}

As described in the Methods, we use two complementary techniques to measure charge noise. For frequencies above 10 mHz we use the Coulomb peak flank method, while for lower frequencies we use Coulomb peak tracking.

The Coulomb flank method follows standard implementations in Refs.~\cite{Connors2019, connors_charge-noise_2022, stehouwerExploitingStrainedEpitaxial2025}. Here we summarize additional implementation details beyond the Methods. The device is biased with a 5 nA low-frequency ac excitation, resulting in an on-chip voltage excitation of $V_S =$ 5 nA$\times$ 25813/$\nu$ $\Omega \approx 43~\upmu$V for $\nu=3$.  Time-series data of the voltage drop $(V_T)$ is recorded at the flank of a Coulomb peak using an SR860 lock-in amplifier after room temperature amplification, and converted to charge noise spectral density $S_\mu$ via Eq.~\ref{eq:smu}, with $G = \nu (e^2/h) V_T/V_S$. To perform wide-bandwidth lock-in measurements, we adjust the time constant of the lock-in such that the 3-dB bandwidth of the low-pass filter after down-conversion filters out the down-converted harmonics of the lock-in frequency, as well as the setup noise above 100 Hz. Additional offline filtering (typically above 20 Hz) is applied during post-processing as described in the Methods. The 3-dB bandwidth of the Gaussian FIR filters in the lock-in amplifier are given by
\begin{equation}
    f_{3\rm{db}} = \frac{1.5}{2\pi t_c}.
    \label{eq:f3dB}
\end{equation}
We use a time constant of 3 ms, corresponding to a 3-dB bandwidth $f_{\text{3dB}} = 80$ Hz. Additionally, we use a lock-in frequency $f_{lo} = 88$ Hz, to ensure that the down-converted second harmonic of the lock-in signal is well outside the 3-dB bandwidth of the lock-in filters. The lock-in amplifier output is recorded in an internal buffer at a sampling rate $f_s = 300$ Hz. We record data on the flank of the Coulomb peaks for a sample time $t_s = 1000$ s. 

After each trace, we re-tune our plunger gate voltage to stay on the flank of a Coulomb peak by performing a single-shot measurement for a range of gate voltages around the previous flank location. We perform these single-shot measurements at a higher time constant of 100 ms, corresponding to a 3-dB bandwidth of $f_{\text{3dB}} = 2.4$ Hz. We wait $t_s = 350$ ms between single-shot measurements for the lock-in to settle, corresponding to a sampling frequency $f_s$ of roughly 3 Hz.

\begin{figure}
    \centering
    \includegraphics[width=\linewidth]{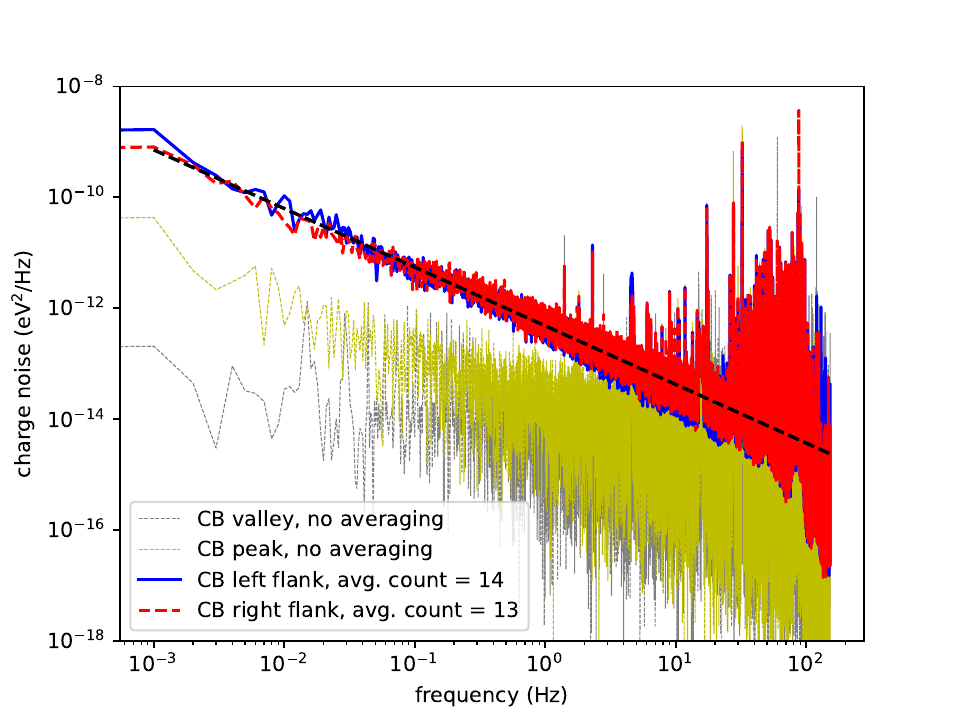}
    \caption{\textbf{Charge noise power spectral density.} Power spectral density curves for time series data taken on the left (blue) and right (red) flanks of a Coulomb blockade peak, as well exactly at the peak position (yellow), and in the valley (gray). The valley and peak noise spectra are not averaged over multiple traces, whereas the Coulomb flank spectra are averaged over the 14 and 13 measured traces for left and right flanks, respectively.}
    \label{fig:coulomb_flank_method}
\end{figure}

To confirm that the measured noise from the Coulomb flank method is dominated by charge noise from the device, we measure the $V_T$ noise both while a) sitting in the valley between Coulomb blockade peaks and while b) sitting on a Coulomb blockade peak. These measurements are done in addition to the noise measurements on the flank of a Coulomb blockade peak, as described in the main text. For charge noise from the device, the measured noise on the flanks should be significantly higher than the noise on the peak or in the valley, since charge noise induces fluctuations in the quantum dot's local chemical potential. We compare the flank data from the main text to measurements from the valley in between Coulomb peaks and from the maxima of the Coulomb peaks themselves, in Fig.~\ref{fig:coulomb_flank_method}. The higher measured noise spectrum for the flanks indicates that the noise is indeed higher than the setup noise floor and is a clear signature of charge noise.

\begin{figure*}
    \centering
    \includegraphics[width=\linewidth]{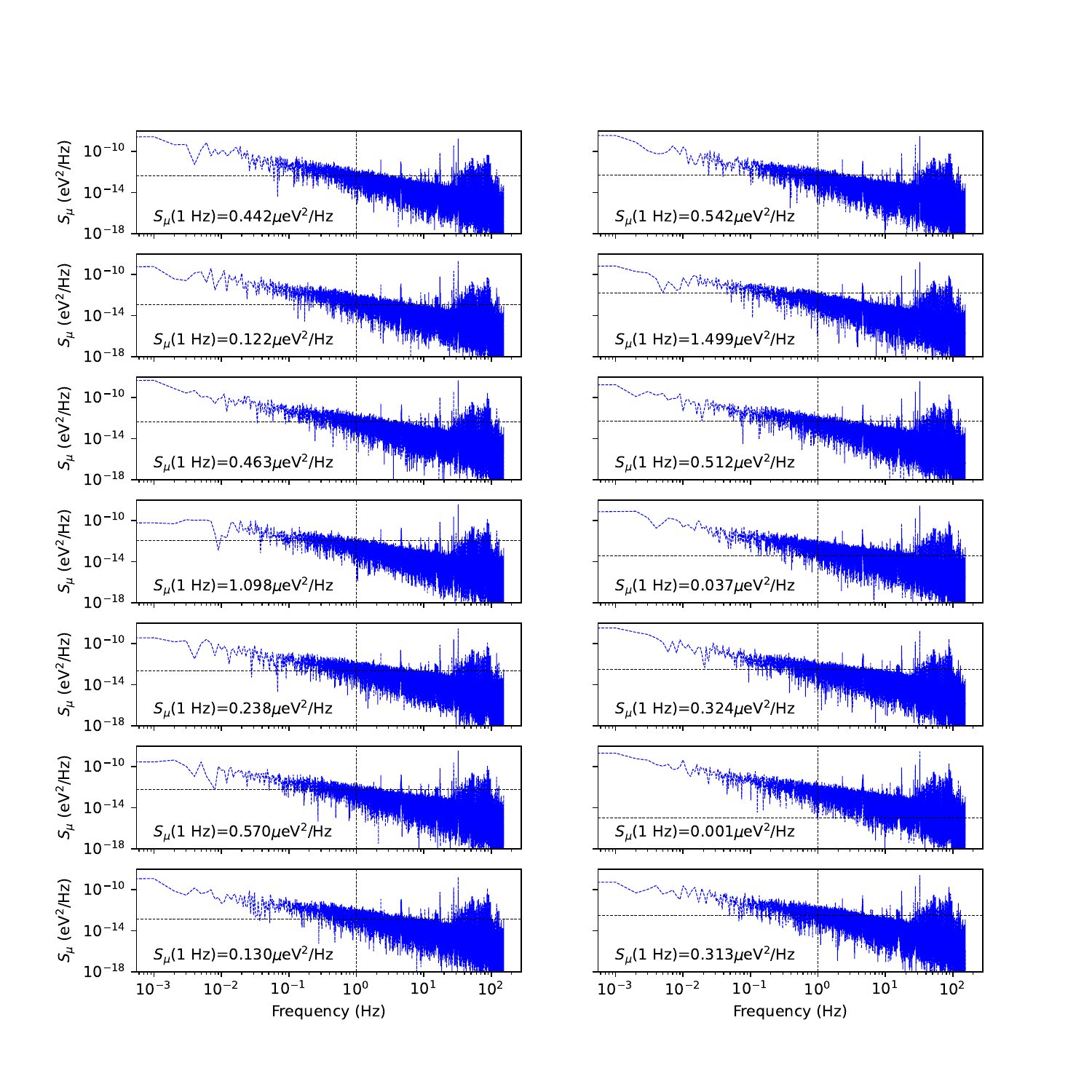}
    \caption{\textbf{Individual charge noise power spectral density traces.}For the Coulomb flank method shown in the main text, the power spectral density curves for each of the 14 peaks that we tracked are plotted, along with a text inset presenting the measured value of the charge noise at 1 Hz, (shown at the intersection of the dashed black lines). Averaging these numbers yields $S_\mu$(1 Hz) = $0.45\pm0.11 \mu$eV$^2$/Hz.}
    \label{fig:individual_psd}
\end{figure*}

Having established that we are measuring charge noise intrinsic to the device, we collect 14 different time-series measurements of the noise using the Coulomb flank method. These 14 traces are converted to power spectral densities of the chemical potential using Eq.~\ref{eq:smu}, and are plotted in Fig.~\ref{fig:individual_psd}. Averaging the measured noise at 1 Hz present in these 14 traces gives our estimated $S_\mu(1\rm{ Hz}) = 0.45 ~\upmu\rm{eV}^2/Hz$.

The second method, Coulomb peak tracking, is used to access lower frequency charge noise where slow drift limits the validity of the linear-response approximation used in the flank method. We sweep the plunger gate voltage in a triangle waveform, back-and-forth across a single Coulomb peak and extract the peak position as a function of time. The resulting time series is made sweep-direction-agnostic to reduce sensitivity to sweep hysteresis. Using the lever arm $\alpha$, this peak-position fluctuation is converted into an equivalent chemical potential noise, from which the low-frequency PSD is computed.

\section{Sensitivity of the conductance to plunger and quantum point contact gate voltages}

\begin{figure}
    \centering
    \includegraphics[width=\linewidth]{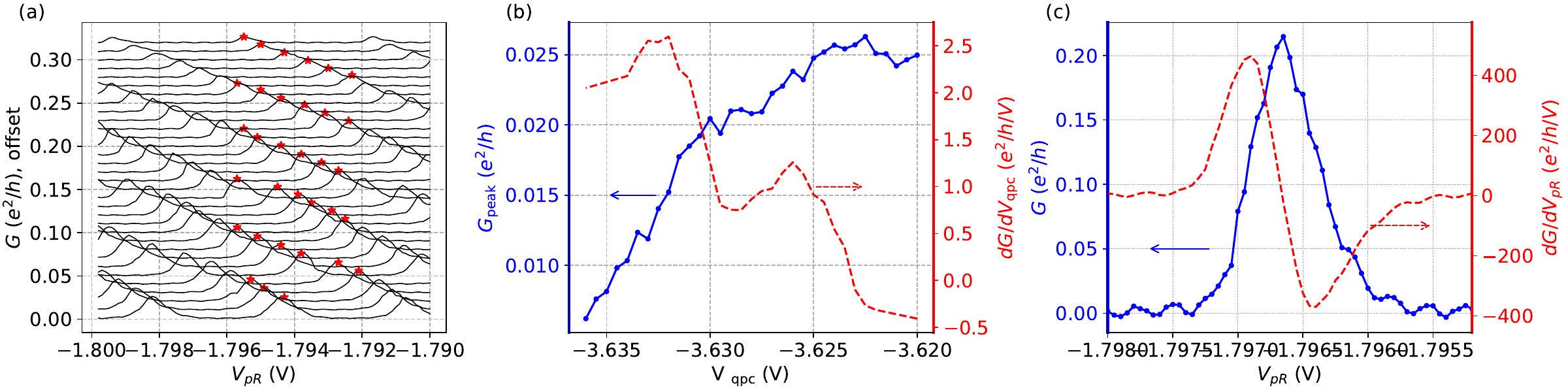}
    \caption{\textbf{Hybrid dot conductance sensitivity to gate voltages.} (a) Device C hybrid dot conductance as a function of the plunger gate voltage $V_{\text{pR}}$. Each trace in (a) corresponds to a fixed voltage $V_{\text{qpc}}\in[-3.62,-3.636]$ V with a step size of $0.5$ mV applied to all the QPC gates. The second conductance peak in each trace is annotated with a red-star. The offset in peak positions is a result of the cross-capacitance between the QPC and the plunger gates. The traces are offset vertically for clarity. (b) The conductance of the second peak for each trace in (a) is plotted as a function of $V_{\text{qpc}}$ (blue), showing an overall decrease in peak $G_{\text{peak}}$ conductance as the transmissions $\tau_{1,2}$ are reduced with $V_{\text{qpc}}$. The red dashed curve shows the slope $dG_{\text{peak}}/V_{\text{qpc}}$. (c) Coulomb peak (blue) for $V_{\text{qpc}} = -3.636$ V and transconductance (red) as a function of plunger gate voltage $V_{\text{pR}}$. As discussed in the main text, the plunger gate is tuned to the flank for CPF measurements, where $dG/d_{V_{\text{pR}}}$ is over 100 times larger than the maximum sensitivity to $dG_{\text{peak}}/V_{\text{qpc}}$.}
    \label{fig:dGdVqpcsmall}
\end{figure}

In this section, we justify the assumption made in the charge noise section of neglecting contributions from fluctuations in the QPC gate voltage $V_{\text{qpc}}$ when evaluating the RMS charge noise in Eq.\ref{eq:rmsmunoise} of the main text.
Figure~\ref{fig:dGdVqpcsmall} presents data from Device C, measured during a separate cooldown, comparing the sensitivity of the hybrid dot conductance to $V_{\text{qpc}}$ and to the plunger gate voltage $V_{\text{pR}}$. A thermally broadened Coulomb peak in the weak tunneling limit is shown in Fig.\ref{fig:dGdVqpcsmall}(c), with the transconductance peaking on the Coulomb peak flank, analogous to Fig.\ref{fig:chargenoise}(b) in the main text.

To assess the effect of small changes in $V_{\text{qpc}}$, we measured a series of Coulomb peaks as a function of $V_{\text{pR}}$, shown in Fig.\ref{fig:dGdVqpcsmall}(a). Each trace corresponds to a different value of $V_{\text{qpc}}$ applied uniformly to all QPC gates, spanning the range $[-3.62, -3.636]$ V in 0.5 mV steps. Due to cross-capacitance between the QPC and plunger gates, the peak positions shift, so we track the peak conductance—defined at $\delta V_{\text{pR}}=0$—as a function of $V_{\text{qpc}}$ in Fig.\ref{fig:dGdVqpcsmall}(b).

The peak conductance values in Fig.\ref{fig:dGdVqpcsmall}(b) are extracted by fitting the annotated Coulomb peaks in Fig.\ref{fig:dGdVqpcsmall}(a) to Eq.\ref{eq:Kulik}. As expected, $G_{\text{peak}}$ decreases as the QPC transmissions $\tau_{1,2}$ are suppressed by more negative $V_{\text{qpc}}$. The slope of this decrease is relatively small, with $|dG_{\text{peak}}/dV_{\text{qpc}}| < 3$ $e^2/h$/V, representing over a 100-fold weaker sensitivity to $V_{\text{qpc}}$ compared to the plunger gate. This is evident in Fig.\ref{fig:dGdVqpcsmall}(c), where the transconductance $dG/dV_{\text{pR}}$ is more than two orders of magnitude larger than the sensitivity to $V_{\text{qpc}}$.

These observations justify neglecting the effect of QPC voltage fluctuations when estimating hybrid dot charge noise.

\section{Charge noise contribution to Coulomb peak width}
In the thermally broadened, weak-coupling limit, the Coulomb peak lineshape for the hybrid dot is described by Eq.~\ref{eq:KSsupp}:
\begin{equation}
G(x) = G_0 \frac{x}{\sinh (x)},
\label{eq:KSsupp}
\end{equation}
where $x = \alpha \delta V_{\text{pR}}/k_BT_e$. Here, $\delta V_{\text{pR}}$ is the plunger gate offset from charge degeneracy, and $T_e$ is the electron temperature. The full width at half maximum (FWHM) of this lineshape, in the absence of charge noise, is given by,
\begin{equation}
\Delta{x}(\sigma_{\upmu} = 0) := \frac{\alpha\left(\delta V_{\text{pR},+} - \delta V_{\text{pR},-}\right)}{k_BT_e} = 4.354,
\end{equation}

where $\delta V_{\text{pR},+}, \delta V_{\text{pR},-}$ are the positive and negative plunger gate voltage offsets from charge degeneracy, respectively, when the conductance is at half its maximum.
Since $\alpha \delta V_{\text{pR}}$ corresponds to the chemical potential $\mu$, we model the effect of charge noise by treating $x$ as a normally distributed random variable, $X \sim \mathcal{N}(x_0, \sigma_x)$, with a plunger gate voltage-dependent mean $x_0 = \alpha \delta V_{pR,0}/k_BT_e$, and standard deviation determined by the total RMS chemical potential noise $\sigma_x = \sigma_{\upmu}/k_BT_e$.% = 2.2 \upmu \text{eV}/4.7 \upmu \text{eV} = 0.47$ at $T_e=54$ mK, based on Eq.~\ref{eq:rmsmunoise}.

As discussed in Eq.~\ref{eq:rmsmunoise}, the total RMS chemical potential noise over the bandwidth corresponding to the measurement time scales is,
\begin{equation}
    \begin{split}
        \sigma_{\upmu} = 2.2~\upmu e\text{V};\implies \sigma_x = \frac{\sigma_{\upmu}}{k_BT_e} = 0.47 
    \end{split}
\end{equation}

The expected peak conductance at $x_0 = 0$ is then given by,
\begin{equation}
\mathbb{E}\left[G_0\frac{X}{\sinh X}\right] = G_0 \int_{-\infty}^{\infty} \frac{x}{\sinh x} \frac{1}{\sqrt{2\pi\sigma_x^2}} e^{-\frac{x^2}{2\sigma_x^2}} dx = 0.965G_0.
\label{eq:xFWHMideal}
\end{equation}

To evaluate the Coulomb peak FWHM in the presence of charge noise, we find the mean offset $X = x^*$ such that $\mathbb{E}[G(X)] = 0.4825G_0$,
\begin{equation}
\mathbb{E}\left[G_0\frac{X}{\sinh X}\right] = G_0 \int_{-\infty}^{\infty} \frac{x}{\sinh x} \frac{1}{\sqrt{2\pi\sigma_x^2}} e^{-\frac{(x - x^*)^2}{2\sigma_x^2}} dx = 0.4825G_0,
\end{equation}
which can be solved numerically, yielding $|x^*| = 2.21$. The FWHM is separation between the two solutions, 
\begin{equation}
x_{\text{FWHM}}(\sigma_{\upmu} = 2.2 \upmu\text{eV}) = 2\times2.21 = 4.42
\label{eq:xFWHMnoise}
\end{equation}

Equations~\ref{eq:xFWHMideal} and~\ref{eq:xFWHMnoise} thus reveal a 1.5\% increase in Coulomb peak FWHM due to chemical potential fluctuations.

\section{Charge quantization in the nearly-ballistic limit}

With elastic cotunneling suppressed by the continuum single-particle spectrum, transport in the hybrid dot is a realization of the inelastic cotunneling model studied by Furusaki and Matveev~\cite{FurusakiMatveev1995}, and is distinct from Yi and Kane's analysis~\cite{YiKane1996} of charge quantization in a phase-coherent quantum dot coupled to leads in the quantum Hall regime. 

When both channels are coupled in the near-ballistic limit, the visibility $Q$ is well described by the asymptotic $\sqrt{1- \tau_2}$ scaling relation,

\begin{equation}
    Q(\sqrt{1-\tau_{1,2}} \ll 1) = \frac{\gamma}{\pi} \frac{E_c}{k_B T_e} \sqrt{(1-\tau_1)(1-\tau_2)},
    \label{eq:Qboson}
\end{equation}
where $\gamma = \exp(C)$, with $C \approx 0.5772$ being Euler's constant, as shown by the dashed curves in Fig.~5(b). The data qualitatively agree with the expected $\sqrt{1-\tau_2}$ asymptotic scaling, with $\tau_1$ left as a fit parameter to account for drift over the course of $G(V_{\text{pR}})$ measurements for a fixed
voltage on the bottom QPC. Deviations
from predictions are likely due to renormalization by dynamical Coulomb blockade, as
discussed in the next section.

A device architecture permitting an independent measurement of $\tau_{1,2}$ would provide a more quantitative analysis of the tunnel couplings, and enable investigations of critical scaling near phase transitions, for example in the multi-channel charge Kondo model ~\cite{FurusakiMatveev1995, Iftikhar2015, Pierre3CK}. In such contexts, the intrinsic, unrenormalized $\tau_{1,2}$ are identified as the relevant microscopic parameters for describing tunnel couplings.

\section{Coulomb diamond and electron thermometry data for Device C}
\label{sec:DeviceC}
\begin{figure*}[!htbp]

     \centering
         \includegraphics[width=1.0\textwidth]{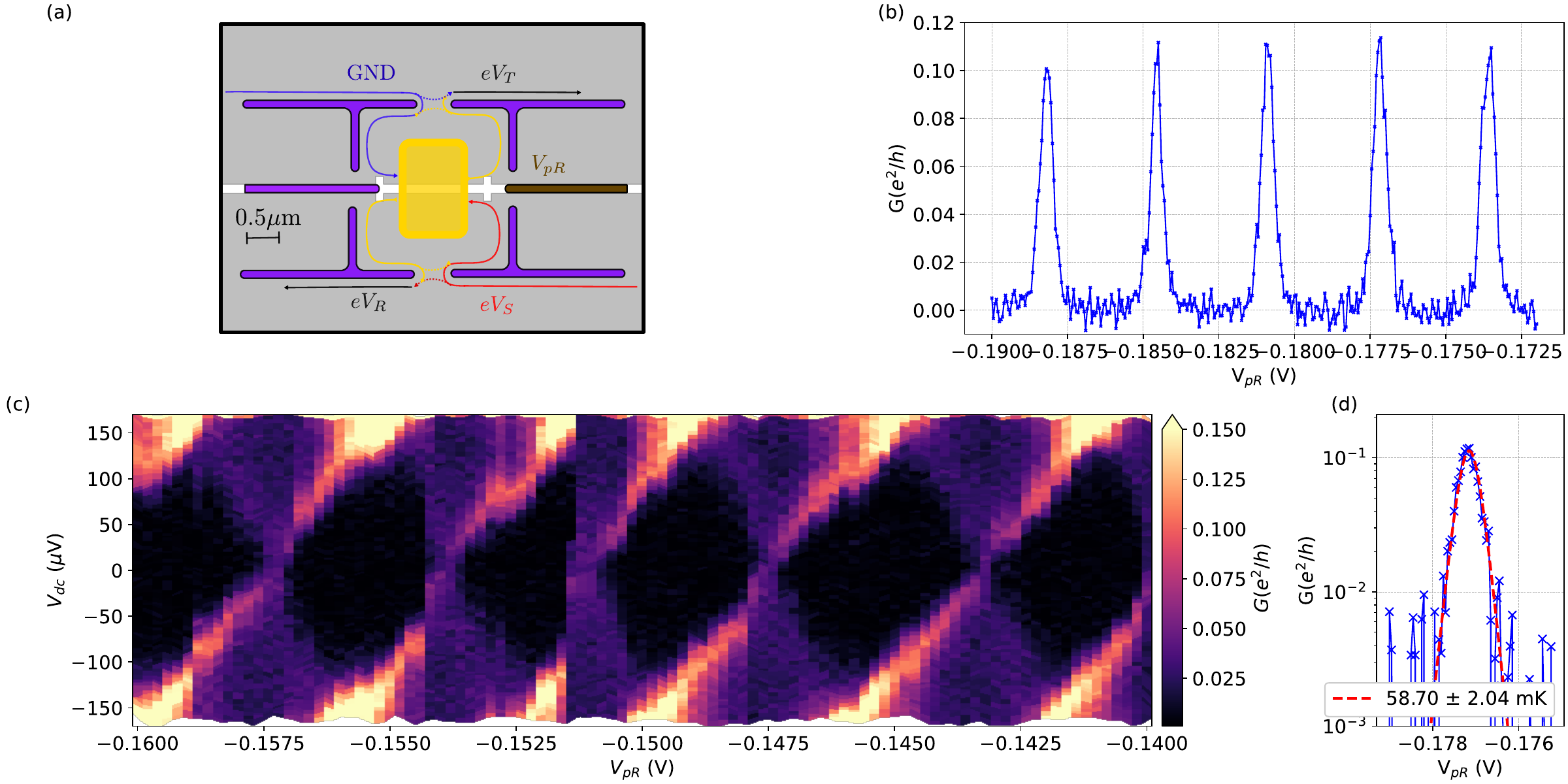}

        \caption{\textbf{Device C Static Coulomb blockade}. (a) Schematic of the hybrid metal-semiconductor
        quantum dot (Device C). A rectangular Ti/Au island with a $0.8\times1.2~\upmu$m$^2$ footprint makes sidewall contact to the InAs quantum well across a 120 nm wide trench (white) on a mesa (gray). The QPC and plunger gates (purple, red) are biased with dc voltage sources, and the right plunger (red) is swept to reveal Coulomb blockade oscillations. The device is biased with a 500 pA low-frequency ($<30$ Hz) ac excitation and a dc current bias $i_{dc}$. A pair of contacts on either end are tied to ground at the mixing chamber stage of the dilution fridge, and realize a ``cold'' ground to drain current. The source ($V_S$), reflected ($V_R$), and transmitted ($V_T$) voltage drops to ground are measured $\sim$ 45-$\upmu$m from the island after amplification by lock-in amplifiers. (b) Coulomb blockade oscillations in the series conductance $G$ as a function of the plunger gate voltage $V_{\text{pR}}$ for a set of 5 successive charging events, with the QPCs tuned to the weak tunneling limit $\tau_{1, 2} \ll 1$. (c) Coulomb diamonds measured by a dc current $|I_{dc}| < 20$ nA. The source dc voltage drop $V_{dc} = I_{dc} \times h/3e^2$ is measured using a multimeter. The charging energy estimated from the diamond heights $E_c = e^2/2C = 70~\upmu eV$, where $C$ is the capacitance of the hybrid dot. (d) A Coulomb peak from (b) on a semi-log scale fit to Eq.~\ref{eq:Kulik} for determining the electron temperature = $59$ mK.}
        \label{fig:DevC}
\end{figure*}

A larger hybrid quantum dot (Device C) fabricated on a different 5$\times$5 mm$^2$ chip was tuned for studying dynamical Coulomb blockade (see Fig.~\ref{fig:DevC}(a)). With the 2DEG tuned to low filling factor quantum Hall, the formation of unintentional quantum dots in a quantum point contact presents a fundamental challenge in realizing smooth and monotonic control of edge mode transmissions. These quantum dots may result from a combination of edge mode reconstruction in a smooth confining potential and disorder-induced density modulation in the incompressible region in the QPC. A QPC design robust to the formation of quantum dots will be addressed in a future work. Although the QPCs in Devices B, and C were nominally identical, the untinentional quantum dots in Device C, if any, were weakly coupled to the edge modes resulting in smooth and monotonic tunability over the entire range shown in Fig.~\ref{fig:DCB}(d). Consequently, Device C presented an attractive playground for a quantitative study of the low bias $eV_{dc} \ll E_c$ suppression of transmission in a resistive environment.

Data for extracting the charging energy $E_c = 70~\upmu$eV for Device C and the electron temperature $T = 59 \pm 2$ mK for the corresponding cooldown is shown in Fig.~\ref{fig:DevC}.

\section{Dynamical Coulomb blockade in the ballistic limit}
\label{sec:DCB}
Although charge quantization is quenched so long as one point contact is in the ballistic limit $\tau_2 = 1$, if the other contact is not in that limit ($\tau_1 < 1$) Coulomb interactions still influence conductance through the dot, suppressing $G$ at low bias. This suppression, known as dynamical Coulomb blockade (DCB), results from the granularity of electronic charge tunneling through the bottom QPC and the finite time constant of the relaxation of electromagnetic modes excited in the resistive environment each time an electron tunnels through this point contact. Considering the bottom QPC, its environment can be modeled as a transmission line with a resistance controlled by transmission through the top QPC ($R_{\text{env}} = (1/\tau_2)h/e^2$), and a capacitance set by the hybrid dot geometry $E_c = e^2/2C$. The tunneling of a single electron through the constriction defined by the bottom QPC displaces the electromagnetic modes in the environment out of equilibrium ~\cite{Flensberg1991}. When $R_{\text{env}} \sim h/e^2$, the time constant for the relaxation of this charge excitation $t_{\text{env}} = R_{\text{env}}C$ is comparable to the time scale of Coulomb interactions $t_{Q} \sim h/E_c$, resulting in charge fluctuations on the order of $e$. For a bias voltage $V_{\text{dc}} < E_c/e$, the dwell time of the electron wavepacket ${t}_{e} \sim h/eV$ exceeds the time scale over which Coulomb interactions set in: ${t}_e > {t}_Q \sim {t}_{\text{env}}$. These Coulomb interactions in the dynamic regime suppress tunneling conductance through the point contact at low bias. For a short conductor in the weak-tunneling limit ($\tau \ll 1$), in the presence of circuit backaction, the suppression in zero-bias transmission was calculated by Joyez and Esteve~~\cite{Joyez1998} and reviewed in Ref.~~\cite{ingold2005charge}. Kindermann and Nazarov~~\cite{KindermannNazarov2003} extended this to arbitrary mesoscopic conductors embedded in a purely dissipative environment, which was generalized to arbitrary environment resistances by Parmentier {\em et al.}~\cite{parmentier_strong_2011}, 

\begin{equation}
    \frac{\tilde{\tau}(V_{\text{dc}})}{1 - \tilde{\tau}(V_{\text{dc}})} = \frac{\tilde{\tau}(V_{\text{dc}}=0)}{1-\tilde{\tau}(V_{\text{dc}}=0)}\frac{E_B(R_{\text{env}}, V_{\text{dc}}, T_e) + 1}{E_B(R_{\text{env}}, V_{\text{dc}}=0, T_e) + 1},
    \label{eq:DCB}
\end{equation}

where $\tilde{\tau}(V_{\text{dc}}) = h/e^2(G^{-1} - R_{\text{env}})^{-1}$ is the transmission extracted from a series conductance measurement at a bias voltage $V_{\text{dc}}$, and $E_B(R_{\text{env}}, V_{\text{dc}}=0, T_e):=\lim_{\tau \to 0} \left(\tilde{\tau}(V_{\text{dc}}=0) - \tau\right)/\tau$ is the relative circuit backaction for a tunnel junction in the same circuit. The backaction $E_B(R, V_{\text{dc}}, T_e)$ for arbitrary bias voltages can be computed using the dynamical Coulomb blockade framework for tunnel junctions (see Methods).

\begin{figure*}
     \centering
         \includegraphics[width=1.0\textwidth]{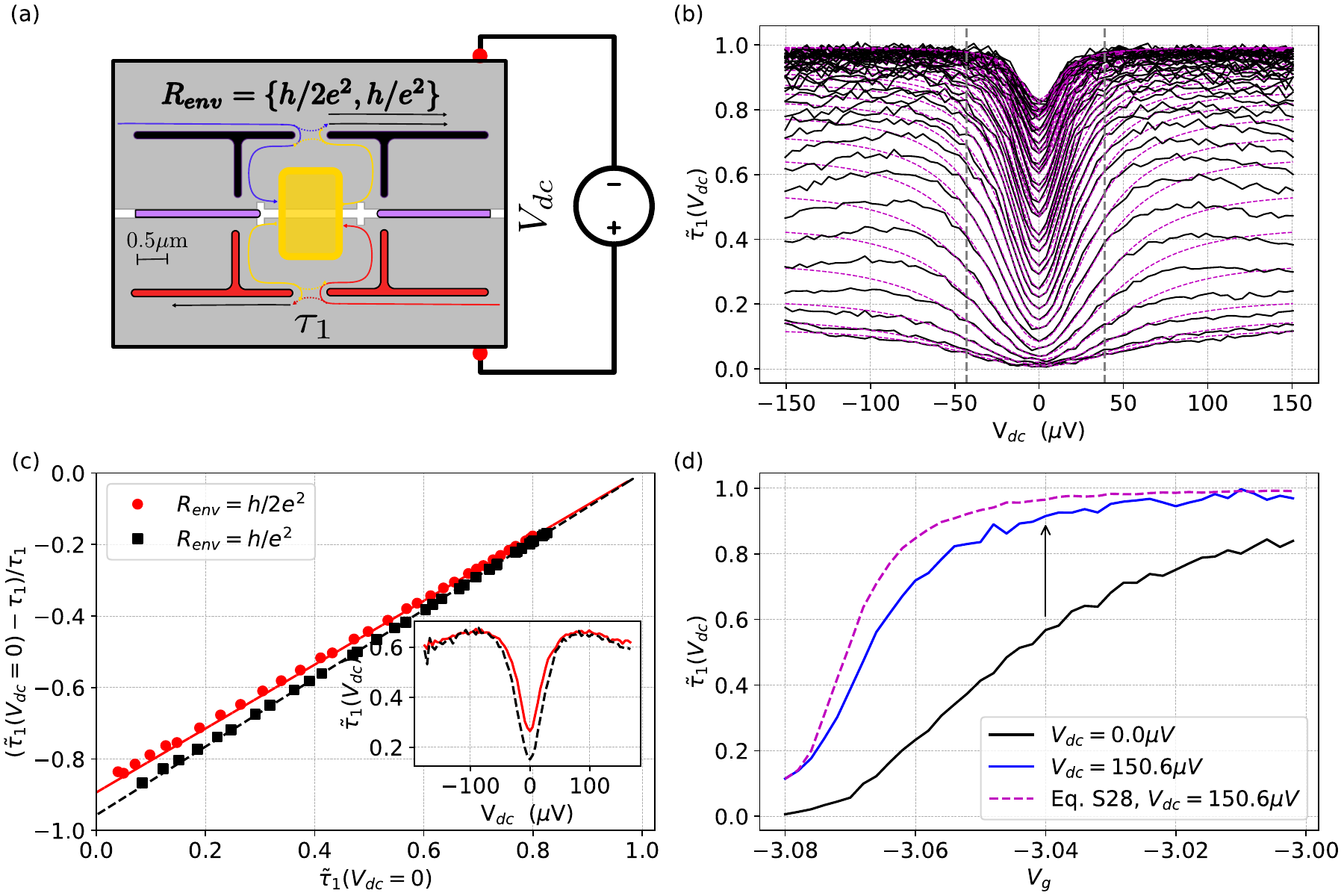}
        \caption{\textbf{Dynamical Coulomb Blockade}. (a) Schematic representation of Device C and the measurement setup. The top QPC is set to transmit one or two edge modes, setting the environment resistance $R_{\text{env}} = h/e^2, h/2e^2$, respectively. A dc current is added to the 500 pA ac current excitation to study the effect of circuit backaction. (b) The bottom QPC gate voltage is stepped up from pinch-off, with the dc bias swept in the range $|eV_{\text{dc}}| < 2E_c$. The low bias $|eV_{\text{dc}}| \ll E_c$ suppression of the transmission extracted from a series conductance measurement $\tilde{\tau}_1(V_{\text{dc}}) = h/e^2\left(G^{-1} - h/e^2\right)^{-1}$ (black traces) is in agreement with DCB predictions (Eq.~\ref{eq:DCB}), shown as pink dashed lines for every QPC gate voltage. The vertical dashed gray lines indicate the dc bias range $|eV_{\text{dc}}| < 43~\upmu\text{eV}$ used for fits to Eq.~\ref{eq:DCB}. (c) The relative circuit backaction as a function of the DCB-suppressed zero-bias transmission, with theoretical prediction for $E_c/k_BT_e = 20$ and $R_{\text{env}} = h/e^2, h/2e^2$. The inset shows dc bias sweeps for the QPC with $R_{\text{env}} = h/e^2, h/2e^2$ at different gate voltages for a $\tilde{\tau}_1 (V_{\text{dc}} = 100~\upmu\text{V)} \approx 0.65$, highlighting the increased backaction in the more resistive environment. (d) Bottom QPC transmission extracted from series conductance measurements plotted as a function of QPC gate voltage at zero dc bias (solid black), and $eV_{\text{dc}} = 150.6~\upmu$V (blue), with the intrinsic low-energy $\tau_1$ prediction from Eq.~\ref{eq:DCB} in dashed pink. The discrepancy between the measured high bias transmission $\tilde{\tau}_1(V_{\text{dc}} = 150.6~\upmu\text{V})$ and the predicted intrinsic low-energy $\tau_1$ is likely due to the energy-dependence of the tunneling through the QPC, and to heating of the dot by applied dc bias.}
        \label{fig:DCB}
\end{figure*}

A larger hybrid dot device (Device C) with $E_c = 70~\upmu$eV, measured in a different cooldown, was tuned for characterizing DCB as depicted in Fig.~\ref{fig:DCB}(a) (see Sec.~\ref{sec:DeviceC} for analogous Coulomb diamond and electron thermometry data). The top QPC is set to transmit one or two edge modes, resulting in an environment resistance $R_{env} = h/e^2, h/2e^2$, respectively.
Figure~\ref{fig:DCB}(b) shows the dc bias-dependence of transmission through the bottom QPC $\tilde{\tau}_1(V_{\text{dc}})$, where each trace corresponds to a particular QPC gate voltage, as the QPC is opened from pinch-off. For a QPC with nearly-energy-independent $\tau$, a bias $eV_{\text{dc}} \gtrsim E_c$ should quench the circuit backaction, so the intrinsic transmission could be read out by measuring at high bias: $\tilde{\tau}_1 (V_{\text{dc}} \gtrsim E_c) \rightarrow \tau_{1}$. However, in our QPCs $\tau_1$ itself has an intrinsic energy dependence, even without DCB. Thus, the high-bias $\tilde{\tau}_1(V_{\text{dc}} \gg E_c)$ is not a simple reflection of the intrinsic low-energy $\tau_1$. Below we describe the modified method we use to determine intrinsic low-energy $\tau_1$, the most fundamental parameter for describing coupling in our system.

At each value of zero-bias transmission $\tilde{\tau}(V_{\text{dc}} = 0)$ set by the bottom QPC gate voltage, the measured bias-dependence of transmission at low bias $(|eV_{\text{dc}}| < 43~\upmu eV)$ is well-described by Eq.~\ref{eq:DCB} (dashed pink traces) based on the known charging energy $E_c$ and the environment impedance $R_{\text{env}} = h/e^2$. The sole fitting parameter $T_e = (40\pm 3)$ mK comes out essentially constant over all values of $\tilde{\tau}(V_{\text{dc}} = 0)$. In contrast, at high bias $eV_{\text{dc}} \geq E_c$ the measured data deviate from Eq.~\ref{eq:DCB}, likely due to the bias dependence of the intrinsic transmission through the saddle-point potential realized by the QPC ~\cite{Hsueh2022}, and to heating by the dc current through the island at high transmissions ~\cite{parmentier_strong_2011}. As noted above, measured conductance at high bias thus does not directly reveal the intrinsic $\tau_1$. Figure~\ref{fig:DCB}(d) shows the transmission through the bottom QPC as a function of QPC gate voltage for $R_{\text{env}} = h/e^2$. Solid curves show transmission extracted from series conductance measurements. Black: $\tilde{\tau}_1(V_{\text{dc}}=0) = h/e^2\left(G^{-1} - R_{\text{env}}\right)^{-1}$ at zero dc bias, strongly suppressed relative to the intrinsic $\tau_1$. Blue: $\tilde{\tau}_1(V_{\text{dc}}=150.6~\upmu\text{V}) = h/e^2\left(G^{-1} - R_{\text{env}}\right)^{-1}$ at $V_{\text{dc}}= 150.6~\upmu$V. Dashed pink curve: low-energy $\tau_1$ extracted from $\tilde{\tau}_1(V_{\text{dc}}=0)$. As described above, at each gate voltage the measured bias-dependent conductance is fit to Eq.~\ref{eq:DCB} for low bias $(|eV_{\text{dc}}| < 43~\upmu\text{eV}$) (Fig.~\ref{fig:DCB}(b)) to extract the low-energy intrinsic $\tau_1$. This would also be the prediction for measured high-bias transmission $\tilde{\tau}_1(eV_{\text{dc}}> E_c)$, if the QPC had no energy-dependence in absence of interactions. 

Equation~\ref{eq:DCB} further predicts a circuit backaction $\left(\tilde{\tau}_1(V_{\text{dc}}=0) - \tau_1\right)/\tau_1$ that is proportional to $(1 - \tilde{\tau}_1\left(V_{\text{dc}}=0)\right)$. This proportionality holds in our measurements, as seen in Fig.~\ref{fig:DCB}(c), which plots the dependence of backaction on the transmission of the bottom QPC. The relevant value of $\tau_1$ is its low-energy limit (dashed pink curve in Fig.~\ref{fig:DCB}(d)), which differs from the measured high-bias $\tilde{\tau}_1$ (blue curve in Fig.~\ref{fig:DCB}(d)). The data plotted as squares and circles are for cases where the top QPC is set to transmit one ($R_{\text{env}} = h/e^2$) or two ($R_{\text{env}} = h/2e^2$) edge modes, respectively. Each of these two data sets shows backaction proportional to $(1 - \tilde{\tau}_1\left(V_{\text{dc}}=0)\right))$, as expected. 

The magnitude of circuit backaction increases as the bottom QPC is pinched off, reaching as much as $80$\% reduction of measured transmission relative to intrinsic transmission. 
The slightly differing slopes of the two data sets match the expectation that the slope should depend on $R_{\text{env}}$. In fact, the two plotted lines which closely match the data sets are not fits, but rather represent the predicted circuit backaction based on Eq.~\ref{eq:DCB} with $E_c/k_BT_e = 20$, $R_{\text{env}} = \{h/e^2, h/2e^2\}$, and no free parameters. The data in the inset show that a more resistive environment produces stronger backaction. The solid trace is for $R_{env} = h/e^2$, whereas the dashed trace is for $R_{env} = h/2e^2$. Each has $\tilde{\tau}_1(V_{\text{dc}} = 100~\upmu\text{V}) \approx 0.65$, as seen in their similar high-bias transmissions, but the low bias transmission is markedly lower for the more resistive environment: $\tilde{\tau}_1(V_{\text{dc}} = 0) \approx 0.15$ compared with $\approx 0.28$.

Our ability to tune $R_{\text{env}}$ and $\tilde{\tau}_{1,2}(V_{\text{dc}})$ enables a systematic study of circuit backaction through the QPC. The lack of a high bias plateau in the measured transmission presents a substantial challenge in calibrating the intrinsic low-energy $\tau_{1,2}$ in the current device architecture. In future designs, incorporating a gate-tunable switch to short-circuit the hybrid dot and quench circuit backaction at zero dc bias ~\cite{parmentier_strong_2011} will make such calibration more reliable.

A device architecture permitting an independent measurement of $\tau_{1,2}$ would provide a more quantitative analysis of the tunnel couplings, and enable investigations of critical scaling near phase transitions, for example in the multi-channel charge Kondo model ~\cite{FurusakiMatveev1995, Iftikhar2015, Pierre3CK}. In such contexts, the intrinsic, unrenormalized $\tau_{1,2}$ are identified as the relevant microscopic parameters for describing tunnel couplings.

\section{Determining the intrinsic transmissions for Device B}

\begin{figure}  
     \centering
         \includegraphics[width=1.0\textwidth]{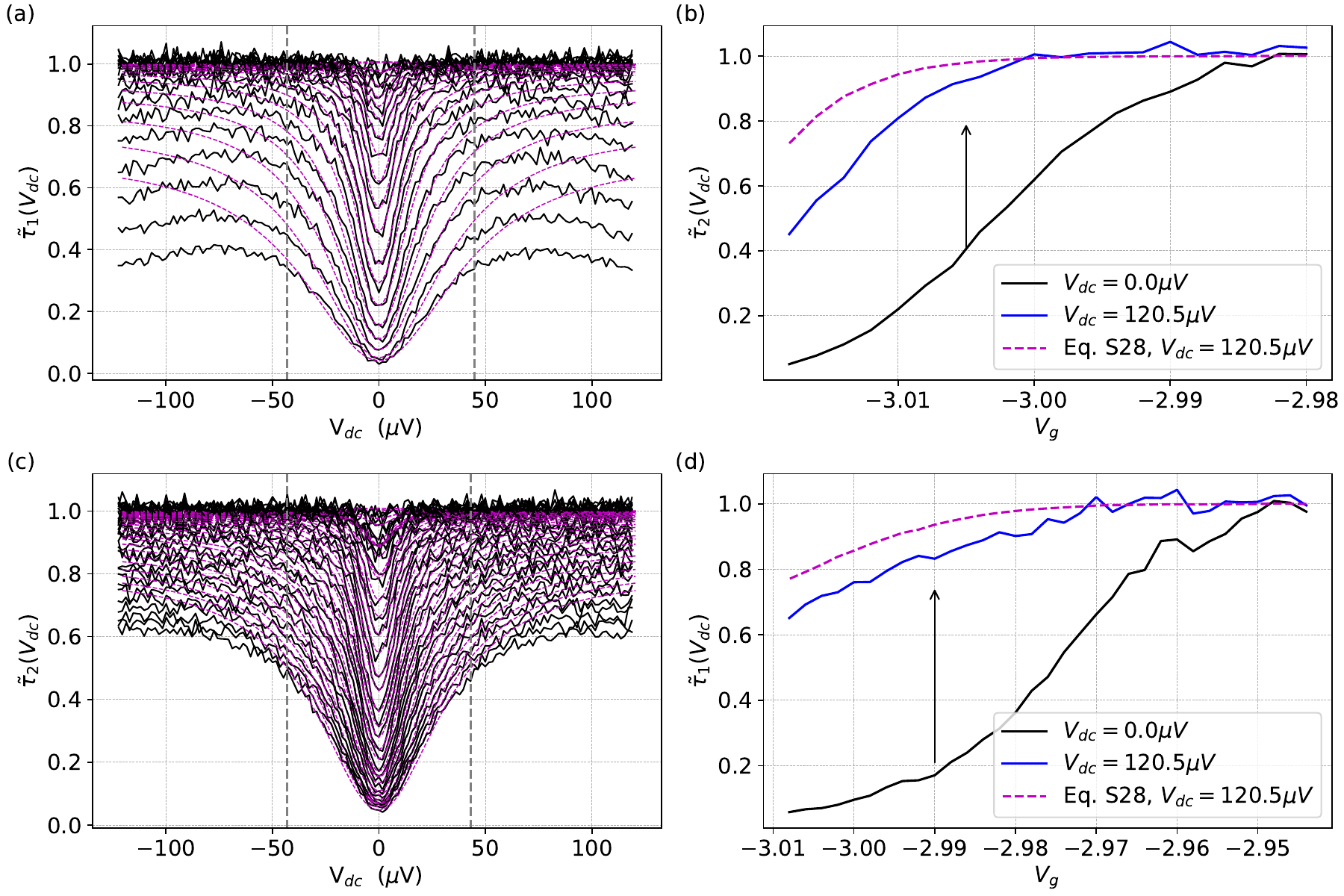}
        \caption{\textbf{Estimating the intrinsic transmission in Device B}. (a) The bias-dependent transmission through the bottom $\tilde{\tau}_1(V_{\text{dc}})$ and (c) top $\tilde{\tau}_2(V_{\text{dc}})$ QPC as the gate voltage is stepped down from the first conductance plateau, with the dc bias swept in the range $|eV_{dc}| < 120.5~\upmu$eV. The low bias $|eV_{dc}| \ll E_c$ suppression of the transmission extracted from series conductance measurements at every gate voltage (black traces) $\tilde{\tau}_{1,2}(V_{\text{dc}}) = h/e^2\left(G^{-1} - h/e^2\right)^{-1}$ is in agreement with DCB predictions (Eq.~\ref{eq:DCB}), shown as pink dashed lines for every QPC gate voltage. The vertical dashed gray lines indicate the dc bias range used for the fits. (b) Bottom and (d) top QPC transmission extracted from series conductance measurements plotted as a function of QPC gate voltage at zero dc bias (solid black), and $eV_{dc} = 120.5~\upmu$V (blue), with the prediction from Eq.~\ref{eq:DCB} in dashed pink. The discrepancy between the measured high bias transmission and the predicted low-energy intrinsic $\tau_{1,2}$ in (b,d) is likely due to the energy-dependence of the tunneling through the QPC.}
        \label{fig:DCB_B}
\end{figure}

The low-energy intrinsic transmissions $\tau_{1,2}$ in Device B are estimated from dynamical Coulomb blockade (DCB) measurements, analogous to the procedure described in Sec.~\ref{sec:DCB}. A dc current $i_{dc}$ is added to the 500 pA ac excitation resulting in a dc bias voltage drop $V_{dc} = i_{dc}\times h/\nu e^2$ for a quantum Hall filling factor $\nu$. The dc bias dependence of transmission through the bottom QPC (top QPC) is extracted from series conductance measurements, with the top QPC (bottom QPC) set to transmit the one edge mode, resulting in an environment impedance $R_{\text{env}} = h/e^2$. Figures~\ref{fig:DCB_B} (a,c) show the dc bias-dependence for the bottom, $\tilde{\tau}_1(V_{dc})$, and top, $\tilde{\tau}_2(V_{dc})$, QPCs respectively, with each trace corresponding to a fixed gate voltage, as the QPC is swept from the first conductance plateau to pinch-off. 

For the zero-bias transmissions $\tilde{\tau}_{1,2}(V_{dc}=0)$ set by the corresponding QPC gate voltages, the measured bias dependence of transmission is fit to Eq.~\ref{eq:DCB} (dashed pink traces) based on the known charging energy $E_c$, and environment impedance $R_{\text{env}} = h/e^2$, with the electron temperature $T_e$ as the sole fitting parameter. The bias range used for the fits in Fig.~\ref{fig:DCB_B}(a,c) are marked by gray dashed lines. The low bias transmission is well-described by Eq.~\ref{eq:DCB}, while at high bias $eV_{dc} \geq E_c$ the measured data in Fig.~\ref{fig:DCB_B}(a,c) deviate from this form.

Figures~\ref{fig:DCB_B}(b,d) show the transmission through the QPCs as a function of QPC gate voltage for $R_{\text{env}} = h/e^2$. Solid curves show transmission extracted from series conductance measurements. Black: $\tilde{\tau}_{1,2}(V_{\text{dc}}=0) = h/e^2\left(G^{-1} - R_{\text{env}}\right)^{-1}$ at zero dc bias, strongly suppressed relative to the intrinsic $\tau_{1,2}$. Blue: $\tilde{\tau}_{1,2}(V_{\text{dc}}=120.5~\upmu\text{V}) = h/e^2\left(G^{-1} - R_{\text{env}}\right)^{-1}$ at $V_{\text{dc}}= 120.5~\upmu$V. Dashed pink curve: low-energy $\tau_{1,2}$ extracted from $\tilde{\tau}_{1,2}(V_{\text{dc}}=0)$. As described above, at each gate voltage the measured bias-dependent conductance is fit to Eq.~\ref{eq:DCB} (Fig.~\ref{fig:DCB_B}(a,c)) to extract the low-energy intrinsic $\tau_{1,2}$. This would also be the prediction for measured high-bias transmission $\tilde{\tau}_{1,2}(eV_{\text{dc}}> E_c)$, if the QPCs had no energy-dependence over the range of applied dc bias, in absence of interactions.

\FloatBarrier

\section{Hamiltonian for a hybrid dot array}
\label{sec:Hamiltonian}
The Hamiltonian for a hybrid dot array linked by chiral quantum Hall edge states may be written as a sum of three components: $H = H_{\text{edge}} + H_{\text{charging}} + H_{\text{QPC}}$. The edge states connecting the metallic islands are described by $H_{\text{edge}} = v_F \sum_{n=1}^{N+1} \int_{-\infty}^{\infty} dx \left[\psi_{L,n}^\dagger i\partial_x\psi_{L,n}-\psi_{R,n}^\dagger i\partial_x\psi_{R,n} \right]$, where $\psi_{\alpha,n}(x)$ annihilates a fermion with chirality $\alpha=L/R$ on the 1D edge connecting islands $n-1$ and $n$; $v_F$ is the Fermi velocity along the edge.
The Coulomb interactions are described by $H_{\text{charging}} = \sum_{\text{island } n} E_c (Q_n - Q_{g,n})^2$. Here, $Q_n=\int_{a}^\infty dx[\psi^\dagger_{L,n}\psi_{L,n}+\psi^\dagger_{R,n}\psi_{R,n}]+\int_{-\infty}^{-a} dx[\psi^\dagger_{L,n+1}\psi_{L,n+1}+\psi^\dagger_{R,n+1}\psi_{R,n+1}]$ denotes the number of electrons on island $n$, while $Q_{g,n}$ is the offset charge controlled by a gate voltage. The charging energy is assumed to be the dominant energy scale. Nevertheless, the charge on the islands is not quantized as long as the connecting channels remain perfectly transmitting. The points $x=-a$ and $x=+a$ are taken to be the boundaries of islands $n$ and $n+1$, respectively, along the line connecting the two islands.
Finally, quasiparticle backscattering at the QPC on link $n$, connecting islands $n-1$ and $n$, is described by the term $H_{\text{QPC}} = \sum_{n=1}^{N+1} r_n(\psi^\dagger_{L,n}\psi_{R,n}+h.c.)$ where $\tau_n \sim 1-|r_n|^2$.

This Hamiltonian exhibits a variety of nontrivial regimes depending on the values of the bare transmission coefficients $\{\tau_n \}$. Setting all backscattering terms to zero, i.e. $\{\tau_n =1 \}$, is expected to yield a metamaterial with nonlocal response properties~\cite{Maillet2026,Sela2026}. Turning on a single reflection coefficient along the chain is predicted to give rise to fractional charge tunneling described by a boundary sine-Gordon model, as previously experimentally simulated for a single island~\cite{Anthore_2018}. For the one-dimensional chain, the effective Luttinger liquid parameter is controlled by the number of islands~\cite{Karki_2022}. For equal reflection amplitudes on two links, not necessarily adjacent, the system is expected to realize a generalization of the Majorana fermions appearing in the two-channel Kondo model: parafermions in the two-site case~\cite{Karki2023}, and more general anyonic excitations with fractionalized statistics for arrays with $N>2$~\cite{Sela2026}. Capacitive interactions between different islands can also be included. 

The most computationally challenging regime corresponds to the strongly interacting case where all inter-dot links have appreciable backscattering. Here, competing contributions in the Hamiltonian generate highly nontrivial correlated behavior. In the limit of strong backscattering between islands, and when the plunger gate voltage is tuned on each island to its charge degeneracy point, the relevant low-energy description is that of chain of charge pseudospins, coupled by \textit{Kondo} rather than simple exchange interactions, as formulated theoretically and demonstrated experimentally for the two-site system in Ref.~\cite{Pouse2023}. When scaled up to a 1D array of many dots, such a setup realizes a kind of Kondo lattice model. Our proposed simulator device in the main text realizes one such system: the SSH-Kondo model. A full analysis of this system and its emergent physics will be presented elsewhere. 

%%%%%%% SUPPLEMENT BIB START %%%%%%%%%%%%%
%apsrev4-2.bst 2019-01-14 (MD) hand-edited version of apsrev4-1.bst
%Control: key (0)
%Control: author (8) initials jnrlst
%Control: editor formatted (1) identically to author
%Control: production of article title (0) allowed
%Control: page (0) single
%Control: year (1) truncated
%Control: production of eprint (0) enabled
\providecommand{\noopsort}[1]{}\providecommand{\singleletter}[1]{#1}%
%

%%%%%%% SUPPLEMENT BIB END %%%%%%%%%%%%%

% \bibliography{aipsamp}
\end{document}